\documentclass[aps,pra,twocolumn]{revtex4-1}
\usepackage{graphicx}
\usepackage{amsmath}\usepackage{commath}
\usepackage{bm}
\usepackage{epstopdf}
\usepackage{amsfonts}
\usepackage{latexsym,enumerate}
\usepackage{epsfig} \usepackage{esvect}
\usepackage[colorlinks=true,linkcolor=blue,urlcolor=blue,citecolor=blue]{hyperref}
\usepackage[T1]{fontenc}
\usepackage[latin9]{inputenc}
\usepackage{hyperref}
\usepackage[bottom]{footmisc}
\hypersetup{
    colorlinks,
    citecolor=red,
    filecolor=green,
    linkcolor=blue,
    urlcolor=brown
}
\hypersetup{linktocpage}
\usepackage{amssymb}
\usepackage{amsthm} \usepackage{mathrsfs}\usepackage{bbm}\usepackage{yfonts}
\newcommand{\ket}[1]{\left\vert#1\right\rangle}
\newcommand{\bra}[1]{\left\langle#1\right\vert} 
\usepackage{upgreek} \usepackage{textgreek} \usepackage{bbold}

\theoremstyle{definition}

\def\beq{\begin{equation}}\def\eeq{\end{equation}}
\def\bearr{\begin{eqnarray}} \def\eearr{\end{eqnarray}}
\def\beal{\begin{equation}\begin{array}{ll}}
\def\eeal{\end{array}\end{equation}}

\begin{document}
\title{ HoloQuantum HyperNetwork Theory }

\author{\vspace{.33 cm} Alireza Tavanfar 
	}
\affiliation{\vspace{.033 cm} Instituto de Telecomunica\c{c}\~{o}es, Physics of Information and Quantum Technologies Group, 
	 Lisbon, Portugal.\\ alireza.tavanfar@cern.ch}
	 \begin{abstract}
The fundamental, general, kinematically-and-dynamically complete quantum many body theory of the entirely-quantized HyperNetworks, namely \emph{HoloQuantum HyperNetwork Theory}, `$\mathcal{M}$', is axiomatically defined and formulated out of a unique system of \emph{nine principles}. HoloQuantum HyperNetworks are all the quantum states of a purely-information-theoretic (0+1) dimensional closed quantum many body system of the abstract qubits of the `absences-or-presences' endowed with a complete distinctive set of abstract many-body interactions among them. All the many-body interactions and the complete total Hamiltonian of $\mathcal{M}$ are uniquely obtained upon realizing the kinematical-and-dynamical quantum-hypergraphical well-definedness by all the `cascade operators', the quantum-hypergraphical isomorphisms, the global-phase $U(1)$ redundancies of the multi-qubit states, the minimally-broken equal treatment of the abstract qubits, the maximal randomness, and `covariant completeness'. Mathematically, HoloQuantum HyperNetworks formulate in completeness all possible unitarily-evolving quantum superpositions of all the arbitrarily-chosen hypergraphs. Physically, HoloQuantum HyperNetworks formulate all the dynamical purely-information-theoretic states of every realizable quantum many body system of the arbitrarily-chosen quantum objects and their arbitrarily-chosen quantum relations. 
Being so, \emph{HoloQuantuam HyperNetwork Theory, $\mathcal{M}$, is proposed as the fundamental, complete and covariant `it-from-qubit' theory of `All Quantum Natures'}.\\ \end{abstract}
\maketitle

\section{
Motives And Methodology}	 

\emph{Hypergraphs} represent \emph{objects and their relations} in the most abstract and general way. The \emph{HoloQuantum HyperNetwork Theory}, `$\mathcal{M}$', which we define, construct and present here upgrades this mere representation to \emph{the form-invariant `it-from-qubit'} \cite{BM16JW} \emph{theory of `all quantum natures'}. That is, all the observables-and-states of all the quantum many body systems of objects-and-relations are covariantly captured by the theory $\mathcal{M}$.\\\\ Mapping time-dependently to the mathematical space of (the maximally oriented and maximally weighted) hypergraphs, the sufficiently-refined choices of the vertices represent the arbitrarily-chosen objects, and the sufficiently-refined choices of the hyperlinks represent the arbitrarily-chosen relations between the objects. Mapping so, the hypergraphical hyperlinks represent all possible physical interactions, classical statistical correlations, quantum entanglement relations, geometric relations, causality relations, structural or functional relations, emergent relations, multi-scale relations, and all possible observable relations. Redefining these hypergraphical representations as \emph{HyperNetworks}, one formulates the dynamical quantum objects-and-relations states-and-observables \emph{from pure information}.\\\\ In quantum realm, the `fundamental information setting' which assembles these objects and relations must be \emph{a special quantum many body system of abstract qubits}. The states of every one of these abstract qubits is spanned on a purely-information-theoretic `no-or-yes', or equally `absence-or-presence' binary basis.\\ Moreover, any closed many body system of the abstract qubits for the quantum objects and their quantum relations undergoes in its totality \emph{a unitary evolution}. The total Hamiltonian of this unitary dynamics must be given by \emph{all the many-body abstract interactions between all these abstract quibits} which are the very fundamental quantum degrees of freedom for all the quantum objects together with all their quantum relations, à la Wheeler \cite{BM16JW}. HoloQuantum HyperNetwork Theory, $\mathcal{M}$, is indeed formulated in this framework. Being so, it formulates all the dynamical mathematically-quantized HyperNetworks whose defining qubits and their abstract interactions all behave as quantumly as in quantum many body systems.\\\\
Being its motive, HoloQuantum HyperNetwork Theory is `The Answer' to the following Question.\\\\
\underline{\emph{Question:}}\\ Let us assume that quantum physics is the exact law-book of the entire universe or the entire multiverse. Let us define \emph{`all quantum natures'} to be the entire quantum universe-or-multiverse, as well as all the arbitrarily-chosen algebraically-closed subsystems of her total observables probed in all the jointly-consistent scales. Equally defined, let all quantum natures be all the theoretically-possible quantum many body systems of quantum objects and their quantum relations. Upon this assumption and  definition, \emph{what is the `it-from-qubit' quantum many body theory which formulates all those quantum natures, both fundamentally and completely, in such a `covariant' way that its quantum kinematics (all the microscopic degrees of freedom) and quantum dynamics (all the many-body interactions and the total Hamiltonian) 
always remain form-invariant?}\\
We will obtain the answer to this question by treating it as an `equation' and then `solving' it constructively.  
The solution must be the unique construction in which two initially-distinguished faces consistently meet one another and merge-in, to form an indistinguishable unit being a `quantum mathematics for all physics'.\\\\ Viewed on its \emph{`mathematics face'}, it must be the fundamental, general and also complete theory of the dynamical entirely-quantized hypergraphs, namely `HoloQuantum Hypergraphs'. Being so, the theory must formulate most fundamentally, and capture most completely, all the dynamical mathematically-quantized hypergraphs which are arbitrarily structured, arbitrarily flavored, arbitrarily weighted and arbitrarily oriented. Viewed on its \emph{`physics face'}, the above theory must be the fundamental formulation of the unitarily-evolving many body system of a complete set of \emph{abstract qubits} which covariantly code the complete time-dependent information of every quantum nature.\\\\ Before undertaking the from-first-principle construction of this unique theory, to be accomplished in sections two to five, we first briefly characterize \emph{the operational way} in which the `all quantum natures' in the original question is being defined. This precise characterization, serving as a preliminary view, will secure that we will be all clear about the defining physics-context of the resulted quantum mathematics.\\\\
\emph{First}, let us consider the \emph{entire quantum universe}, or to be maximally general (as we must), the \emph{entire quantum multiverse}. Physically this quantum universe or quantum multiverse is defined as the maximally-large maximally-complete set of $C^\star$-algebra-forming quantum observables which are based on the Hilbert space of the total set of independent quantum degrees of freedom. By construction, the evolving global wavefunctions of \emph{this initial all-inclusive quantum many body system} contain the complete time-dependent information which are decodable by all possible measurements on it.\\\\
\emph{Second}, we now consider all possible operations of \emph{`phenomenological reductioning'}. By definition, every such operation corresponds to an arbitrarily-chosen (proper) subset of the entire set of quantum observables on the Hilbert space of an arbitrarily-chosen (proper or improper) subset of that total set quantum degrees of freedom, which are scale-preserved in the sense of the renormalization group flow, and all together form a  $C^\star$-subalgebra. Every such phenomenological reduction, being a likewise-reduced quantum many body system, observationally defines \emph{a `phenomenologically-reduced quantum universe or multiverse'}. Clearly now, for every reduced universe-or-multiverse, the likewise-constructed evolving global wavefunctions contain the complete time-dependent information which are decodable by all possible measurements on it.\\ 
\emph{Third}, to every such collection of the $C^\star$-algebra-forming quantum observables and their base Hilbert spaces, being the complete universe-or-multiverse or each one of the reduced universes-or-multiverses, we apply all possible operations of \emph{`probes rescalings'}, by which one's scale of probing the observables is being re-chosen in all the arbitrarily-chosen ways (that are only constrained by all the algebraic-and-observational consistencies). Let us call the resulted quantum many body systems the \emph{`renormalized quantum universes-or-multiverses'}. Now, all as before, the likewise-constructed evolving global wavefunctions of every one of these renormalized quantum universe-or-multiverse contains its complete time-dependent information.\\\\
\emph{`A quantum nature'}, as we name, is every so-defined quantum universe-or-multiverse, being the entire one, a reduced one, or a renormalized one. As every universe is indeed a physically-realizable quantum many body system, we choose the initial system to be `the ultimate quantum multiverse', defined as the set-theoretic union of all the in-principle-realizable universal quantum many body systems. Being so, it must be now manifest that \emph{`all quantum natures'} are certainly re-identifable as all the `theoretically-possible' quantum many body systems of the arbitrarily-chosen quantum objects and their arbitrarily-chosen quantum relations.\\ 
\\ As a theory of all physics, what this work is all about, is to construct from first principles the covariant total state-space and the covariant total Hamiltonian of the most-fundamental purely-information-theoretic quantum many body theory whose unitarily-evolving quantum states encode the complete information of every such-framed quantum nature. The HoloQuantum HyperNetwork Theory which we present in this work perfectly accomplishes \emph{this 
`it-from-qubit'  
mission}. 
\section{HoloQuantum HyperNetwork 
Theory:
The Threefold Foundation\\ And The Quantum Kinematics}
HoloQuantum HyperNetwork Theory is axiomatically defined and formulated based upon nine principles. In constructing a theory, the choice of its principles is almost everything, so we highlight a cardinal point about these nine principles. To the best of our understanding, the chosen principles are the ones which serve the theory as the unavoidable, and at the same time, as the very best foundational statements to fulfill its intentional mission, as was stated and explained in section one. Namely, \emph{these nine principles are `the must be' for consistency, and `the best be' for the fundamental-ness}.\\\\
We now begin to answer the original question by stating the three principles which set the \emph{`threefold foundation' of HoloQuantum HyperNetwork Theory}.\\
{\bf Principle 1:} HoloQuantum HyperNetwork Theory, `$\mathcal{M}$', must be axiomatically defined and constructed to be the kinematically-and-dynamically covariant complete formulation of all the in-principle-realizable quantum many body systems whose defining degrees of freedom are the arbitrarily-chosen quantum objects and their arbitrarily-chosen quantum relations, capturing perfectly all their time-dependent states and observables. Stated equally, it must be the fundamental complete theory of  `all quantum natures'. Mathematically, to accomplish the above mission, the theory $\mathcal{M}$ must be the general complete theory of HoloQuantum Hypergraphs, namely, the hypergraphs whose all-quantized defining structures become the quantum degrees of freedom constituting an interacting quantum many body system which evolves unitarily.\\\\ 
\emph{Explanatory Note}: We must highlight that, both the objects and their relations must be taken-in as the original notions and the quantum degrees of freedom of the presented theory. Because all relations, by their very conception, are to be defined between some objects, the theory, in order to be both the most fundamental and the most complete, does also need to have the quantum objects included in the totality of its microscopic degrees of freedom. But hierarchically viewed, relations can also be `objectified' and so then develop the relations between relations. Being so, upon straightforward hierarchical extensions which keep the formulation of the theory invariant, the defining quantum degrees of freedom will additionally include all these \emph{`hierarchical relational relations'}. Because these hierarchical extensions are all straightforward, they are left implicit in the present work.\\
\\  {\bf Principle 2:} To accomplish the mission specified in principle one, $\mathcal{M}$ must be an entirely-pregeometric quantum field theory, defined and formulated in $(0+1)$ spacetime dimensions. Namely, it must be `a' theory of quantum mechanics.
\\\\ \emph{Explanatory Note}: HoloQuantum HyperNetwork Theory, at the level of its fundamental formulation, lives in $(0+1)$ spacetime dimensions. However, its sub-theories, model extractions, solutions and phases realize all the in-principle-realizable many body systems of quantum objects and their quantum relations defined in arbitrary numbers of spatial dimensions.\\
\\ {\bf Principle 3:} The theory $\mathcal{M}$ must be the `it-from-qubit' fundamental theory of all quantum natures. Being so, HoloQuantum Hypergraphs must be formulated purely information-theoretically as the unitarily-evolving states of an unprecedented quantum many body system of `abstract qubits', namely HoloQuantum HyperNetworks. By definition, each one of these fundamental qubits represents the states of the `absence-or-presence' of a corresponding (sufficiently-refined) quantum vertex or a (sufficiently-refined) quantum hyperlink.
\\ \emph{Explanatory Note}: `Formally', the abstract qubits of $\mathcal{M}$ shall be formulated by means of fermionic operators. That is, the binary basis-states of each one qubit are represented by the eigenvalues of the number operator of a representing fermion, being either zero or one. Clearly, as an `it-from-qubit' theory, the final computations of all the true observables of $\mathcal{M}$ are invariant under how one formulates its abstract qubits. Being so, these same qubits can be equally represented by spins or by any other alternatives. But, as the `it-from-qubit' theory of all quantum natures, $\mathcal{M}$ embodies its simplest first-rate formulation in terms of the \emph{fermionically-represented qubits}. Highlighting that this choice is simply a very convenient formality to represent our abstract qubits, let us name them \emph{`formally-fermion qubits'}.\\\\ We now merge these three principles into one unit: \\\\ 
\underline{\emph{The Threefold Foundation of} $\mathcal{M}$:}\\ $\mathcal{M}$ will be formulated as `a' closed quantum many body system of interacting (formally-fermion) abstract qubits of the absences-or-presences of the arbitrarily-chosen quantum objects and their arbitrarily-chosen quantum relations, living in $(0+1)$ dimensions. The global wavefunctions of all these qubits must correspond in a one-to-one way to the unitarily-evolving quantum superpositions of the arbitrarily-structured hypergraphs.\\\\ To formulate HoloQuantum HyperNetwork Theory, let us choose an arbitrary $M \in \mathbb{N}$, and let $\mathcal{H}_{M}^{(\mathcal{M})}$ \emph{be the defining total Hilbert space of all the HoloQuantum Hypergraphs whose structures can take not more than $M$ present vertices}. The degrees of freedom of $\mathcal{H}_{M}^{(\mathcal{M})}$ are a total number of $M^\star = M^\star(M)$ formally-fermion qubits of the `absences or presences', each one for a quantum vertex or for a quantum hyperlink. We take a finite $M$, but are also free to take $M \to \infty$. Moreover, let us introduce an arbitrarily-chosen all-index `$\;I\;$' which \emph{even-handedly} but uniquely labels all the $M^\star$ qubits as $F_{I}$. With every one qubit $F_I$ come its creation and annihilation operators $(f_I,f_I^\dagger)$, forming the algebra,
				\begin{equation}\label{frmalgbra} 
				f^{\dagger 2}_I = f_J^2 = 0 \;\;\;;\;\;\; \{f_J,f^\dagger_I\} = \delta_{JI} \:\:\:,\;\;\;\forall\;I\;,\;\forall\;J
				\end{equation} 
We name the \emph{qubit} for a \emph{vertex} a `\emph{verton}', and the \emph{qubit} for a \emph{hyperlink} whose identification needs `$m$' number of defining `base vertices' an \emph{`$m$-relaton'}, where $m \in \mathbb{N}^{\leq M}$.
Because mathematically, a hypergraph is defined free from any background space,
 $\mathcal{M}$ must be fundamentally defined in zero space dimensions. Being so, at the level of the fundamental formulation of the theory, vertons 
are labeled with one single index $i \in \mathbb{N}^{\leq M}$ to address their independent identities. We denote all the vertons of $\mathcal{M}$ by $V_i$, together with their fundamental operators $(v_i,v^\dagger_i)$. Accordingly, the $m$-relatons of $\mathcal{M}$
are addressed uniquely as $R_{i_{(m)}}$, and $(r_{i_{(m)}},r^\dagger_{i_{(m)}})$, with $i_{(m)} \equiv \{i_1, ..., i_m\}$ being their \emph{sets} of base-vertons. Upon \emph{any} one-to-one map, %
\begin{equation} \{{\rm all}\;i^{\in \mathbb{N}^{\leq M}}\} \;\cup \; \{ {\rm all}\; i_{(m \in \mathbb{N}^{\leq M})}\} \equiv\; \{I\}^{{\rm Cardinality} \;=\; M^\star} \end{equation} 
Let us highlight a clarifying point. In this section, given its mission, we only need to formulate \emph{`the basic class' of HoloQuantum HyperNetworks}. By definition, these are HoloQuantum HyperNetworks for which the relatons are totally insensitive to the ordering of their base vertons, and moreover, the relatons and the vertons are weightless and flavorless. But, these basic-class minimalizations will be all lifted up where \emph{in section five}, on the basis of the results in sections two to four, \emph{we will perfect both the mathematical and physical generality of $\mathcal{M}$ by formulating the HoloQuantum HyperNetworks whose corresponding hypergraphs are `maximally flavored', `maximally weighted', and also `maximally oriented'}. Maximal orientedness means that the base-ordering degeneracy of the relatons are entirely lifted up, so that every relaton is defined by the unique ordering of all its base vertons.\\\\ 
Now, the fundamental operators of the formally-fermion qubits of $\mathcal{M}$ are so collected, 
\begin{equation}
\begin{split} 
&\{\;f^{(\dagger)}_I\,;\;|\{I\}| = M^\star
\;;\;M^\star = 2^M-1+M\; \} \; \equiv \;\\&\equiv \{v^{(\dagger)}_{i}\; ;\; r^{(\dagger)}_{i_{(1)}} \:,\; r^{(\dagger)}_{i_{(2)}}\; ,\; r^{(\dagger)}_{i_{(3)}}\; , \cdot \cdot \cdot ,\; r^{(\dagger)}_{i_{(M)}} |\;{\rm all\;possible\;indices}\} 
\end{split}
\end{equation} 
The \emph{simplest observables} of the 
theory are the number operators of all the vertons and all the relatons, 
\begin{equation}  n_{I} \equiv f^\dagger_I f_I\;\;\;;\;\;\;n_I^2 = n_I \;\;\;;\;\;\;[n_I,n_J] = 0 \;\;\;,\;\;\;\forall\;I\;,\; \forall\;J  \end{equation}
Multi-qubit states are defined as the common eigenstates of the $n_I$ observables. 
Given the standard vacuum state, 
\begin{equation}\label{svs} 
\ket{0} \in \mathcal{H}_{M}^{(\mathcal{M})}\;\;\;;\;\;\; f_I \ket{0} = 0\;\;\;,\;\;\forall\; I \end{equation} 
the `\emph{$m$-qubit states}' are built as follows,
\begin{equation}\label{eqa5}  
\prod^{1\leq s \leq m}_{I_s} f^\dagger_{I_s}\ket{0}\;,\:\forall\; \{I_s\}_{{|}_{1 \leq s \leq M}}^{\subset \{I\}^{{\rm Cardinality\;} M^\star}}\;\;,\;\forall\;m \in \mathbb{N}^{\leq M^\star}
\end{equation}
But, the defining state-space of the theory $\mathcal{M}$, namely $\mathcal{H}_{M}^{(\mathcal{M})}$, is much smaller than the space spanned by all the multi-qubit wavefunctions constructed in (\ref{eqa5}).
This must be clear because, as one demands, \emph{for a multi-qubit wavefunction in (\ref{eqa5}) to be a defining basis-state of} $\mathcal{H}_{M}^{(\mathcal{M})}$, \emph{it must be identifiable with a mathematically-welldefined `classical' hypergraph}. Restated, each multi-qubit state in (\ref{eqa5}) belongs to the defining basis of $\mathcal{H}_{M}^{(\mathcal{M})}$, if for every relaton $R_{i_{(m)}}$ being present in it, all of its base-vertons $\{V_{i_{1}}, \cdot \cdot \cdot , V_{i_{m}}\}$ are also present in it. This condition, which discerns the basis of $\mathcal{H}_{M}^{(\mathcal{M})}$, mathematically represents the statement that the presence of a relation without the presence of every one of its identifying objects render its containing wavefunctions meaningless. 
This point yields the \emph{`consistency quantum constraints'} whose \emph{kinematical version} comes in this section, but their \emph{dynamical version} will come in section three.\\
For every qubit $F_I$, let us introduce its uniquely-defined  `qubit-presence projection operator', being denoted by $P_I$. By definition, it is the unique operator by which any superposition of the states in (\ref{eqa5}) is projected down to the maximal part of it in which every participating state has the presence of the qubit $F_I$. By fermionic algebra,
\begin{equation}\label{projectors} P_I\; =\; n_I \:\:\:;\:\: \bar{P}_I \;\equiv\; 1 - P_I\; =\; 1 - n_I \;\;\;,\;\;\;\forall\; I \end{equation}
Given (\ref{projectors}), let us express here the \emph{`kinematical quantum constraints'} that discern which ones among the $m$-qubit states in (\ref{eqa5}) form the basis of $\mathcal{H}_{M}^{(\mathcal{M})}$, namely $\mathcal{B}_{M}^{(\mathcal{M})}$,
		 		\begin{equation}\begin{split}\label{equa6a}
		 		& \forall \;{\rm state}\;\ket{\hat{\Psi}}\;{\rm in} \;(\ref{eqa5})\;\;\;,\;\;\;\ket{\hat{\Psi}}\;\in \mathcal{B}_{M}^{(\mathcal{M})}\;\;\;;\;\;{\rm if}:\\& Y_{i_{(m)}} \ket{\hat{\Psi}} \;\equiv\;\; n_{i_{(m)}}\; \big( 1 \;- \prod_{\;{\rm all}\;i_s\;\in\; i_{(m)}} n_{i_s} \big) \ket{\hat{\Psi}}  = 0 \;\;\;,\;\;\forall\;R_{i_{(m)}} 	 
		 	\end{split}
		 	\end{equation}
		 	Constraints (\ref{equa6a}) are solved by replacing the $r^{(\dagger)}_{i_{(m)}}$s, one by one, with their `vertonically-dressed' counterparts $\breve{r}^{(\dagger)}_{i_{(m)}}$s,	        \begin{equation}\label{relatons} \hspace{.15 cm}
             \breve{r}^{(\dagger)}_{i_{(m)}} \;\equiv\:\; r^{(\dagger)}_{i_{(m)}} \prod_{\;{\rm all}\;i_s\;\in\; i_{(m)}} n_{i_s}\;\;\;;\;\;\;\breve{n}_{i_{(m)}}\;=\;n_{i_{(m)}} \prod_{\;{\rm all}\;i_s\;\in\; i_{(m)}} n_{i_s}   
            \end{equation}
            By definition (\ref{relatons}), it must be clear that the operators (\ref{relatons}) bring about only the \emph{`hypergraph-state relatons'}, namely the `correct relatons'.
Keeping vertons intact, we now evenhandedly define the \emph{`hypergraph-state-qubits'} $\breve{F}_{I}$ as,
            \begin{equation}\label{aub}       \breve{f}_I \;\equiv\; f_I \prod_{{\rm all\;the\;base}\;i_s\;{\rm for}\; I} n_{i_s} \;;\;\; \breve{f}^\dagger_I
            \;\equiv \prod_{\;{\rm all\;the\;base}\;i_s\;{\rm for }\; I} n_{i_s}\;\; f^\dagger_I\;\;\;,\;\;\forall\;I
            \end{equation}
            Now,  \emph{the total Hilbert space of} $\mathcal{M}$, \emph{namely} $\mathcal{H}_{M}^{(\mathcal{M})}$, \emph{must be defined by its most-complete vertonic-and-relatonic tensor-product structure} \emph{and its so-identified basis} $\mathcal{B}_M^{(\mathcal{M})}$, 
             \begin{equation}\label{equa7}\begin{split}
             &  \mathcal{H}_{M}^{(\mathcal{M})}\; =\; \mathcal{H}_{(\rm{vertons})}^{(\mathcal{M})}\; \otimes_{{}_{1 \leq m \leq M}}\; \mathcal{H}_{(m- {\rm relatons})}^{(\mathcal{M})} \\             
            & \mathcal{B}_{M}^{(\mathcal{M})} \;=\; 
            \big\{\; \ket{I_1 \cdot \cdot \cdot I_m}\; ,\;\forall\;\{I_s\}_{|_{1 \leq s \leq m}}^{\subset \{I\}^{{\rm Cardinality}\;M^\star}},\;\forall\; m \in \mathbb{N}^{\leq M^\star} \big\}\\ & \hspace{1.66 cm} 
             \hspace{.11 cm} ({\rm where}:) \hspace{.35 cm} \ket{I_1 \cdot \cdot \cdot I_m}\; \equiv\; \ket{\vec{I}_{\;\norm{m}}}\;\equiv \prod^{1\leq s \leq m}_{I_s} \breve{f}^{\dagger}_{I_s}\;\ket{0} 
 \end{split}\end{equation} 
By the second and the third lines of (\ref{equa7}), the basis $\mathcal{B}_M^{(\mathcal{M})}$ is defined to be the set of the common-eigenstates of all the number operators $n_I$s in which a possible collection of the hypergraph-state-qubits $\{\breve{F}_{I_1} \cdot \cdot \cdot \breve{F}_{I_m}\}$ are present, while all the other hypergraph-state-qubits $\{ \breve{F}_I \}^{({\rm Cardinality}\;M^\star)} - \{\breve{F}_{I_1} \cdot \cdot \cdot \breve{F}_{I_m}\}$ are absent. Being so, every $\ket{\vec{I}_{\;\norm{m}}}^{\in\;\mathcal{B}_M^{(\mathcal{M})}}$ corresponds to a unique `classical' hypergraph, and so identifies the set of all $\ket{\vec{I^\prime}_{\;\norm{m}}}^{\in\;B_M^{(\mathcal{M})}}$ which as classical hypergraphs are isomorphic to it. We will elaborate on this classification in principle five.\\ \emph{Result} (\ref{equa7}), \emph{concludes the quantum kinematics of} $\mathcal{M}$. But, let us highlight two points. 
            \emph{First}, HoloQuantum HyperNetwork Theory is, and must be, a `fully-internal theory', with no `external' qubits in it whatsoever. All its qubits  correspond \emph{information-theoretically} to the mathematical structures of HoloQuantum Hypergraphs. Vertons are qubits \emph{for} objects, the vertices, and relatons are qubits \emph{for} relations, the hyperlinks. Clearly, there can be no other qubits in an `it-from-qubit' theory of all quantum natures. \emph{Second,} HoloQuantum HyperNetwork Theory is the theory \emph{`of'} the mathematically-quantized hypergraphs, not a quantum theory \emph{`on'} hypergraphs. All possible backgrounds and excitons appear only as sub-theories and models, solutions and phases of $\mathcal{M}$.\\\\ As of now, our task is developing axiomatically all the many-body interactions, and the total Hamiltonian of HoloQuantum HyperNetworks, which determine both their complete unitary dynamics and phase diagram.
\section{HoloQuantum HyperNetwork Theory: 
The Fundamental Rule\\ 
And Cascade Operators}
Let us now begin to work out \emph{the dynamical side} of $\mathcal{M}$. In this section, based on the next five principles, principles four to eight, we formulate \emph{`the fundamental rule' of} $\mathcal{M}$, namely, its core many-body interactions.\\\\  HoloQuantum HyperNetwork Theory, formulated  fundamentally, must be the theory of \emph{closed} quantum many body system of all the vertons and all the relatons. This `closed-system criterion', which is already imposed by principle one, comes on two regards. \emph{Firstly}, for being the quantum many body theory of all quantum natures, the above criterion must be met necessarily, because every quantum universe-or-multiverse must be a closed many body system. \emph{Secondly}, this is also a prerequisite to $\mathcal{M}$ to be able to formulate all forms of HoloQuantum HyperNetworks, fulfilling its mathematical completeness. Because, having got the theory of `closed HoloQuantum HyperNetworks', `open HoloQuantum HyperNetworks' are formulated by any partitioning of the total system of vertons-and-relatons, and deriving the master equation out of the total unitary evolution in $\mathcal{M}$.
Being a closed quantum system of all $\breve{F}_I$ s, the total evolution of the HoloQuantum HyperNetworks must be \emph{this unitary one},
 \begin{equation}\label{equa8}
             	U_M^{({\mathcal{M})}}(t) \equiv 
             	e^{i t H_M^{(\mathcal{M})}}\;;\; H_M^{(\mathcal{M})} \equiv\; \sum^{{\rm all}}_{\upkappa}\; \lambda_{\upkappa}\;
             	\mathcal{O}^{(H^{(\mathcal{M})}_M)}_\upkappa
             	[\{\breve{f}_J\}\;;\;\{\breve{f}_I^\dagger\}] 
             \end{equation}
             By the right-hand side of (\ref{equa8}), the \emph{total Hamiltonian} of $\mathcal{M}$, $H_M^{({\mathcal{M})}}$, must be a sum over \emph{all the `Hamiltonian Operators'} $\mathcal{O}^{(H^{(\mathcal{M})}_M)}_\upkappa$ which, as composite operators made from the $f_J$s and the $f^{\dagger}_I$s, realize all the many-body interactions between the vertons-and-relatons.\\
We will \emph{uniquely} determine (\ref{equa8}), \emph{by realizing the nine principles}. Going forward, we set the next four principles.\\\\ {\bf Principle 4:} The `hypergraph-state well-definedness' of HoloQuantum HyperNetworks must be preserved during their entire unitary evolutions whose Hamiltonian, in its maximally-general form, is expressed as (\ref{equa8}).\\\\ 
    \emph{Explanatory Note}: Remember that HoloQuantum HyperNetworks are the unitarily-evolving superpositions of the $\mathcal{B}_{M}^{(\mathcal{M})}$-states given in (\ref{equa7}). \emph{Being  central to} $\mathcal{M}$, principle four demands that, when $U_M^{({\mathcal{M})}}(t)$ in (\ref{equa8}) acts on each basis-state in (\ref{equa7}) chosen as an initial state, the `vertonically-unsupported relatons' should not take presence in the wavefunctions. The resulted \emph{dynamical quantum constraints} will be solved by a whole family of \emph{`Cascade Operators'}.
\\\\ {\bf Principle 5:} In HoloQuantum HyperNetwork Theory, all the operators in $\mathcal{S}^{(\mathcal{M})}_M$, which realize (the second quantized form of) all the hypergraphical isomorphisms, must be made out of some of the exact Hamiltonian operators $\mathcal{O}^{(H^{(\mathcal{M})}_M)}_{\upkappa}$ introduced in (\ref{equa8}). By definition, every HoloQuantum HyperNetwork is transformed by these operators into all possible states which are quantum-hypergraphically isomorphic to it, in such a way that its basis-states (\ref{equa7}) (as `classical' hypergraphs) are either kept invariant automorphically or transformed isomorphically. Being so constructed, these operators must then complete the purely-vertonic conversions by realizing all the relatonic adjustments induced by them.\\
\\\emph{Explanatory Note:} $\mathcal{M}$ calls for the above principle because of being mathematically \emph{`the quantum theory of hypergraphs'}. Let us define how the elements of $\mathcal{S}^{(\mathcal{M})}_M$ must be constructed. First, one defines an abstract discrete space $\mathcal{V}^{\mathcal{(M)}}_M$, the `vertonic field space', for which the indices of all the $M$ vertonic degrees of freedom of $\mathcal{H}_{M}^{(\mathcal{M})}$ form a `coordinate system'. Now, every element of $\mathcal{S}^{(\mathcal{M})}_M$ is a 
permutational coordinate transformation on $\mathcal{V}_M^{(\mathcal{M})}$, $\{V_i\} \to \{V_{\mathcal{P}(i)}\}$, being accompanied by all the `induced' adjustments on the indices of relatons, $\{R_{i_{(m)}}\} \to \{ R_{\mathcal{P}(i)_{(m)}} \}$. Realized as field operations in the second-quantized framework, such transformations are all realized by the specific products of the so-constructed \emph{`order-one isomorphism operators'},
$\Gamma_j^i$, 
\begin{equation}\label{11vcp}\begin{split} & \; \Gamma_{j}^i\; \equiv\;  \mathbb{1}\; \bar{P}_j \;+\;  [\; \mathbb{1}\; P_i\; +\;  (v_j \mathcal{G}_j^i v^\dagger_i)\; \bar{P}_i\; ]\; P_j\; =\\ &  \hspace{.00533 cm} =\;\;1 + n_j(1 - n_i)\;(v_j \mathcal{G}_j^i v^\dagger_i - 1)\;\;\;,\;\forall\; (j , i)  \end{split} \end{equation} which act on HoloQuantum HyperNetworks $\ket{\Psi}^{\in\;\mathcal{H}_M^{(\mathcal{M})}}$. According to (\ref{11vcp}), the permutation `$j \to i$' is relazied by the purely-vertonic operator $v_j v_i^\dagger$ and is simultaneously completed as an isomorphism by the purely-relatonic operator $\mathcal{G}_j^i$, fulfilling the relaton adjustments as needed by the vertonic inductions.\\ 
{\bf Principle 6:} The global $U(1)$ transformations of the hypergraph-state-qubit operators $f^{(\dagger)}_I$ which generate global phases on the multi-qubit states of $\mathcal{B}_M^\mathcal{(M)}$ in (\ref{equa7}) should be all redundant observationally. Being so, these transformations form a quantum-mechanical global $U(1)$ fundamental symmetry of $\mathcal{M}$ under which all the observables, states, many-body interactions, and the unitary evolutions of HoloQuantum HyperNetworks must be invariant.
\\\\\emph{Explanatory Note:} HoloQuantum HyperNetwork Theory is an inherently background-less quantum field theory. As such, considering the theory at its fundamental level, there can be no spatial-background potential or non-trivial spatial-background topology by which the global phases of the multi-qubit wavefunctions can become observables. On the other hand, by the identification of $\mathcal{B}_M^{(\mathcal{M})}$ in (\ref{equa7}), these redundant phases of the basis-states are all generated by the \emph{global} $U(1)$ transformations on the fundamental operators of the hyper-state-qubits $f^{(\dagger)}_I$s, which so must be the global $U(1)$ symmetry transofrmations of $\mathcal{M}$.
	\\\\ {\bf Principle 7:} All the relatons, independent of both their defining hyperlink degrees and their base vertons, must be treated equally in the fundamental formulation of HoloQuantum HyperNetwork Theory. Moreover, to incorporate the vertons and the relatons all together, the theory must treat all the hypergraph-state-qubits $\breve{F}_I$ with the `maximum-possible level of equality', at the level of its fundamental formulation. That is, this `all-qubits equal-treatment' will be explicitly but only minimally broken by merely the kinematical and dynamical quantum constraints for the hypergraph-state dependencies of the relatons to their base vertons. Nevertheless, by this minimally-broken symmetry, the theory $\mathcal{M}$ is as maximally HyperNetworkical as it can be.
\\\\\emph{Explanatory Note:} Realizing this principle, the quality of hypergraph-ness is maximally realized in $\mathcal{M}$, both kinematically and dynamically. Being one of its most distinctive characteristics, this principle is `a must' for $\mathcal{M}$ to be the fundamental complete covariant theory of all quantum natures. Specially, this cardinal feature becomes manifested in three ways. \emph{First}, In $\mathcal{M}$, the higher-degree relatons are elementary and so \emph{irreducible} degrees of freedom which are not composites of any lower-degree relatons. In particular, the two-relatons are neither the `building-block relatons', nor by any other means `the more special' relatons. \emph{Second,} all the $m$-relatons must participate in the many-body interactions and in the total Hamiltonian of $\mathcal{M}$ in an equal manner. \emph{Third}, there indeed must be interactions in $\mathcal{M}$ which convert the $m$-relatons and the vertons into one another, in all possible ways which are consistent with the dynamical hypergraphical well-definedness of HoloQuantum HyperNetworks and with other principles.\\ 
{\bf Principle 8:} $\mathcal{M}$, to be the fundamental and complete quantum many body theory of all quantum natures, must be not only a `fundamentally-random theory' but also a `maximally-random theory'. Namely, at the level of the fundamental formulation of $\mathcal{M}$, absolutely all the many-body interactions must be defined up to the maximally-random ensembles of their defining couplings. \\\\\emph{Explanatory Note:} $\mathcal{M}$, by its original intention of being the most fundamental theory of all quantum natures, must be constructed in \emph{`the maximal resonance'} with the statistical randomness which is at the centre of the \emph{principle of `Law Without Law' by Wheeler} \cite{BM16JW}. This also resembles the randomness in the matrix-and-tensor models \cite{BM1WTPRG}. In HoloQuantum HyperNetwork Theory, this fundamental randomness is `maximally' realized. That is, $H_M^{(\mathcal{M})}$ receives all the many-body-interaction operators $\mathcal{O}_{\upkappa}^{(H_M^{(\mathcal{M})})}$ by the couplings $\lambda_{\upkappa}$ whose statistical distributions are `maximally random'. We choose here to take all the couplings to be (the real-or-complex valued) continuous parameters, although they may be also taken to be parameters with any discrete spectra. By this choice, \emph{the maximally-free Gaussian distributions} for all the $\lambda_k$ serve the formulation of  $\mathcal{M}$ as the most natural ones. Principle eight supports the `embedding completeness' of the theory additionally, because then all the deterministic theories are also embeddable in it, by suitably fixing the two characteristic parameters of the corresponding maximally-free Gaussian distributions.\\\\
Already before the step-by-step construction of $U_M^{({\mathcal{M})}}(t)$, we state \emph{The Fundamental Rule of $\mathcal{M}$} which later in this section is obtained by \emph{merging principles one to eight}.\\\\ \underline{\emph{The Fundamental Rule of $\mathcal{M}$}:}\\ By the  first-order many-body interactions of $\mathcal{M}$, namely by its core microscopic operations, every arbitrary pair of the hypergragh-state-qubits $(\breve{F_J},\breve{F}_I)$, either one being any verton or any $m$-relaton, can become converted to one another, $\breve{F}_J \leftrightarrows \breve{F}_I$. The corresponding strengths of these core operations are set by the maximally-free Gaussian-random couplings $(\bar{\lambda}_J^I, \lambda_J^I)$. If the conversion is purely relatonic, $\breve{r}_{j_{(m_j)}} \to \breve{r}_{i_{(m_i)}}$, no accompanying conversions are required. But, if the conversion pair contains a verton, $v_j \leftrightarrows \breve{f}_I$, a cascade of  
purely-relatonic conversions will be jointly triggered. These companion conversions, which secure the dynamical hypergraphical well-definedness of HoloQuantum HyperNetworks, are realized by the `First-Order Cascade Operators' $(\mathcal{C}_j^{I \dagger},\mathcal{C}_j^I)$.\\\\ 
Now, by the realization of all the above eight (out of nine) principles, we systematically construct the first-order many-body interactions of $\mathcal{M}$, namely the above fundamental rule. But also, \emph{the correct forms} of the first-order cascade operators and the first-order Hamiltonian of $\mathcal{M}$ are determined. This shows the road map to construct $U_M^{(\mathcal{M})}(t)$ by realizing principle nine. \\ Let us begin this mission with \emph{realizing principle four}.
One must let the initial HoloQuantum HyperNetwork be an arbitrary quantum state on the basis (\ref{equa7}). Then the initial wavefunction (or mixed state) clearly satisfies all the kinematical constraints (\ref{equa7}), by being a superposition of (or a density operator made from) the basis-states which are all the well-defined classical hypergraphs. Now, principle four is the demand that the total HoloQuantum HyperNetwork at any time $t$, evolved by $U_M^{(\mathcal{M})}(t)$ in (\ref{equa8}) from any such initial wavefunction, must be also a superposition of the very same basis-states (\ref{equa7}). On one hand, this demand simply states that the Hamiltonian whose most general expression is given in (\ref{equa8}), $H^{(\mathcal{M})}_{M} [ \{\breve{f}_J\}_{{|}_{{}_{\forall J}}} ; \{\breve{f}_I^\dagger\}_{{|}_{{}_{\forall I}}} ]$, must serve $\mathcal{M}$ as an operator which is consistently defined on the constrained state-space (\ref{equa7}). That is, its `basis-dependent expression' must be of the form, 
\begin{equation}\label{hta} 
H^{(\mathcal{M})}_{M}[ \{\breve{f}_J\}_{{|}_{{}_{\forall J}}} ; \{\breve{f}_I^\dagger\}_{{|}_{{}_{\forall I}}} ] = \sum_{\vec{J}_{\norm{s}},{\vec{I}_{\norm{r}}}} h({\vec{J}_{\norm{s}}}|{\vec{I}_{\norm{r}}}) \ket{\vec{I}_{\norm{r}}} \bra{\vec{J}_{\norm{s}}} 
\end{equation}
On the other hand, we need to obtain the alternative form of $H_M^{(\mathcal{M})}$ which, \emph{first of all}, manifestly satisfies all the nine principles, and \emph{second of all}, is basis-independently determined as a composite operator which is directly, solely and explicitly made from the `alphabet'  operators $\{\breve{f}_J\}_{{|}_{{}_{\forall J}}} \cup \{\breve{f}_I^\dagger\}_{{|}_{{}_{\forall I}}}$, as meant by the right hand side of (\ref{equa8}).
Aiming so, let us now work out the explicit statement of the principle four. As being stated, all the HoloQauntum HyperNetworks, evolved from the initial superpositions of the basis-states (\ref{equa7}) by the unitary evolution operator (\ref{equa8}), should also satisfy (\ref{equa6a}). That is,
\begin{equation}\label{equa88}\begin{split}
&\forall\; \ket{\Psi}\;\;\; {\rm satisfying}\;\;\; Y_{i_{(m)}}\ket{\Psi} = 0\;\;,\;\forall\; R_{i_{(m)}}\\&{\rm also\;must\;have}:\; Y_{i_{(m)}}\;\big( U_M^{({\mathcal{M})}}(t)\ket{\Psi} \big) = 0\;\;,\;\;\forall\; R_{i_{(m)}}\;,\;\forall\;t
\end{split}\end{equation}
Because $U_M^{({\mathcal{M})}}(t)$ is generated by $H_M^{(\mathcal{M})}$, the conditions (\ref{equa88}) can be equally restated as the following conditions,
\begin{equation}\label{8ab} 
[H^{(\mathcal{M})}_M , Y_{i_{(m)}}]\; \propto \; {\rm function\;of\;the\;constraints}\;Y_{j_{(s)}}\;,\;\;\forall\; R_{i_{(m)}}
\end{equation}
Every operator $\mathcal{O}^{(H_M^{(\mathcal{M})})}_\upkappa \equiv \mathcal{O}^{(H_M^{(\mathcal{M})})}_\upkappa [\{\breve{f}_J\} ; \{\breve{f}_I^\dagger\} ]$ in (\ref{equa8}) is an individual term of the total Hamiltonian. Being so, because of (\ref{8ab}), it must also satisfy the constraints,  
\begin{equation}\label{8abc} 
[\mathcal{O}^{(H_M^{(\mathcal{M})})} , Y_{i_{(m)}}] \propto \; {\rm function\;of\;the\;constraints}\;Y_{j_{(s)}}\;, \forall\; R_{i_{(m)}}
\end{equation}
We call the above quantum constraints (\ref{8ab}), or equally (\ref{8abc}), the \emph{`dynamical quantum constraints' of $\mathcal{M}$}. Now, for the general HoloQuantum HyperNetwork to satisfy principle four, \emph{we directly solve the quantum constraints} (\ref{8abc}). Besides the important simplest solutions to (\ref{8abc}), we obtain \emph{the equally-important `cascade-operator solutions' of the one-to-one conversions}, to be later generalized by means of the `descendant cascade operators'.\\ Clearly enough, the most straightforward family of the general solutions to 
(\ref{8abc}) are given by \emph{the purely-relatonic many-body operators}, namely by the following operators,
\begin{equation}
\mathcal{O}^{({\rm purely\;relatonic})}_{\{j_{a^\prime (s_{a^\prime})}\},\{i_{a (r_a)}\}} \equiv \big( \prod_{1 \leq a^\prime \leq m^\prime} \breve{r}_{j_{a^\prime (s_{a^\prime})}} \big) \big( \prod_{1 \leq a \leq m} \breve{r}^\dagger_{i_{a (r_a)}} \big)
\end{equation}
The complementary class of solutions to (\ref{8abc}) will be given by \emph{the many-body operators which contain in them an arbitrary number of the verton operators}.
Such a vertonically-involved many-body operator generically does not satisfy all the dynamical constraints in (\ref{8abc}). This 
must be clear because, if a vertonic annihilation operator $v_j$ acts on a basis-state in (\ref{equa7}), generically it turns it into a hypergraphically-incorrect multi-qubit wavefunction, as all the $V_j$-based relatons which were present in the at-the-time state of the system, would be now leftover with incomplete vertonic support. Hence, every such $v_j$-action must be accompanied by a cascade operator whose defining action protect the resulted state against the vertonically-unsupported relatons. Here we obtain the \emph{first-order cascade operators}.\\\\ First, let us consider the operator $\mathcal{O}^{(H_M^{(\mathcal{M})})}$ by which a verton $V_j$ is converted to a verton $V_i$. Because the simplest option $v_j v_i^\dagger $ can not be a Hamiltonian operator, 
we must enlarge it by a yet-unknown \emph{cascade operator} $\mathcal{C}_j^i$, and demand that the operator $v_j \mathcal{C}_j^i v^\dagger_i$ satisfies (\ref{8abc}). We now determine the \emph{the correct form of the cascade operator} $\mathcal{C}_j^i$. To fulfill its intented mission, $\mathcal{C}_j^i$ must be divisible into the following product form,
\begin{equation}\label{18}
\mathcal{C}_j^i \;=\;\prod_{m = 1}^{M}\; \prod_{y_{(m)}\; \equiv\; \{y_1 \cdot \cdot \cdot y_m\}}^{{\rm all\;choices}}\;\mathcal{C}^{j \to i}_{y_{(m)}}
\end{equation}
Every contributing $y_{(m)}$ in the right-hand side of the above product (\ref{18}), identifies an arbitrary choice of $m$ vertons $\{V_{y_1}, \cdot \cdot \cdot, V_{y_m}\}$ out of the $M-2$ available vertons setting aside $\{V_i , V_j\}$. In (\ref{18}) all such choices of $y_{(m)}$s are incorporated impartially. Moreover, all the internal operators $\mathcal{C}^{j \to i}_{y_{(m)}}$ must be expandable in this form,
\begin{equation}\label{gbegin}
\begin{split}
\mathcal{C}^{j \to i}_{y_{(m)}}\; =\; A_{\perp}^{\perp}\;\bar{\breve{P}}_{i y_{(m)}} \bar{\breve{P}}_{y_{(m)} j}\; +\;  A_{\perp}^{\parallel}\;\bar{\breve{P}}_{i y_{(m)}} \breve{P}_{y_{(m)} j} \\ +\;\;\;A_{\parallel}^{\perp}\;\breve{P}_{i y_{(m)}} \bar{\breve{P}}_{y_{(m)} j} \;+\; A_{\parallel}^{\parallel}\;\breve{P}_{iy_{(m)}} \breve{P}_{y_{(m)} j} 
\end{split}  
\end{equation}
In (\ref{gbegin}), the indices `$y_{(m)} j$' and `$i y_{(m)}$' identify the two $(m+1)$ relatons $\breve{R}_{y_{(m)} j}$ and $\breve{R}_{i y_{(m)}}$ which relate the $m$ vertons $\{V_{y_1}, \cdot \cdot \cdot, V_{y_m}\}$ to $V_j$ and $V_i$, respectively. Also, 
$(\breve{P}_{y_{(m)} j},\bar{\breve{P}}_{y_{(m)} j})$ and 
$(\breve{P}_{i y_{(m)}},\bar{\breve{P}}_{i y_{(m)}})$ are the doublets of the projection operators which respectively correspond to the presences and the absences of $\breve{R}_{y_{(m)} j}$ and $\breve{R}_{ iy_{(m)}}$.
Furthermore, there are two sets of operator identities that all the $\mathcal{C}_j^i$s, and so as induced by them, each one of the $\mathcal{C}^{j \to i}_{y_{(m)}}$ s, must satisfy. These identities are as follows,
\begin{equation}\label{sina1}
 (\mathcal{C}_j^i)^\dagger = \mathcal{C}_i^j \; \Rightarrow\; \big(\mathcal{C}^{j \to i}_{y_{(m)}}\big)^\dagger = \mathcal{C}^{i \to j}_{y_{(m)}} \;;\;\;\; C_j^j = 1\; \Rightarrow\; \mathcal{C}^{j \to j}_{y_{(m)}} = 1 \;
\end{equation}
By the left-side of (\ref{sina1}), the two inversely-conversional operators $\mathcal{O}^{(H_M^{(\mathcal{M})}) i}_j \equiv v_j \mathcal{C}_j^i v^{\dagger}_i$ and $\mathcal{O}^{(H_M^{(\mathcal{M})}) j}_i \equiv  v_i \mathcal{C}_i^j v^{\dagger}_j $ add up to 
a Hermitian unit of 
$H^{(\mathcal{M})}_M$. Right-side condition in (\ref{sina1}), on the other hand, must be required for the following `interaction-consistency criterion' to be realized. The number operator $n_j = v_j^\dagger v_j $, which surely is a Hamiltonian operator by satisfying (\ref{8abc}), can be regarded as the conversion of $V_j$ into the same $V_j$, and so can be likewise recast as $n_j = 1 - v_j \mathcal{C}_j^j v_j^\dagger$, yielding $\mathcal{C}_j^j = 1$.\\\\
By applying the two sets of the operator identities in (\ref{sina1}) to the general expansion 
in (\ref{gbegin}), one obtains,
\begin{equation} 
\label{gmiddel}\begin{split}
& \mathcal{C}^{j \to i}_{y_{(m)}}\; =\; \big( \bar{\breve{P}}_{iy_{(m)}} \bar{\breve{P}}_{y_{(m)}j} + \breve{P}_{iy_{(m)}} \breve{P}_{y_{(m)}j} \big) 
\\ 
&\hspace{.85 cm} + \;\; A_{\perp}^{\parallel} \; \bar{\breve{P}}_{iy_{(m)}} \breve{P}_{y_{(m)}j} + \; (A_{\perp}^{\parallel})^\dagger \; \breve{P}_{iy_{(m)}} \bar{\breve{P}}_{y_{(m) }j} 
\end{split} \end{equation}   
Now, for the $V_j \leftrightarrows V_i$ conversions to realize \emph{principle four}, one demands that the enlarged operators $(v_j \mathcal{C}_j^i v^\dagger_i , v_i \mathcal{C}_i^j v^\dagger_j)$ satisfy the dynamical constraints (\ref{8abc}). By satisfying this requirement together with realizing \emph{principle five} in the corresponding form which realizes (\ref{11vcp}), and finally upon utilizing $(\breve{P}_I,\bar{\breve{P}}_I) = (\breve{n}_I,1-\breve{n}_I)$, we conclude as follows \emph{the purely-vertonic first-order cascade operators $\mathcal{C}_j^i$}, \begin{equation}\label{cor}\begin{split}
& \mathcal{C}_j^i \;=\; \prod_{m =1}^M\; \prod_{y_{(m)}}^{{\rm all\;choices}}\;\mathcal{C}^{j \to i}_{y_{(m)}}\\
& \mathcal{C}^{j \to i}_{y_{(m)}}\; =\; \big(1 - \breve{n}_{iy_{(m)}} - \breve{n}_{y_{(m)}j} + 2 \:\breve{n}_{iy_{(m)}} \breve{n}_{y_{(m)}j}\big)
\\
& \hspace{.66 cm}\hspace{.22 cm} +\; (1-\delta_{ij})\;\big(\;\breve{r}^\dagger_{y_{(m)}j}\;\breve{r}_{iy_{(m)}}\; +\; \breve{r}^\dagger_{iy_{(m)}}\;\breve{r}_{y_{(m)}j}\; \big)
\end{split} \end{equation} 
Now, using fermionic algebra, and because the cascade operator $C_i^j$ always appears in $H_M^{(\mathcal{M})}$ in the form $v_j C_j^i v_i^\dagger$, one clearly sees that \emph{the cascade operators} (\ref{cor}), 
\emph{can be also operationally simplified in the following form}, 
\begin{equation}\label{sco}
\mathcal{C}_j^{i}\; =\; \prod_{m=1}^{M}
\prod_{y_{(m)}}^{{\rm all\;choices}}\;\big[\;\big(1\; -\; \breve{n}_{y_{(m)}j}\big)\;
+\; \breve{r}^\dagger_{iy_{(m)}}\;\breve{r}_{y_{(m)}j}\;\big] 
\end{equation}
We now move on to formulate the \emph{first-order interactions which realize th conversions between vertons and relatons} in HoloQuantum HyperNetwork Theory. As above, it will be done by the Hamiltonian operator $v_j \mathcal{C}_j^I \breve{f}_I^\dagger$ which converts a verton $V_j$ to \emph{any hypergraph-state-qubit} $\breve{F}_I$. By \emph{realizing principle seven}, and by (\ref{sco}), $\mathcal{C}_{j}^{I}$ must be so,
\begin{equation}
\mathcal{C}_{j}^{I} \equiv\; \mathcal{C}_j^k\; [\;{\rm if\;index}\;k\;{\rm being\;replaced\;with\;the}\;{\rm index}\;I\;]\;,\;\forall\;I \hspace{.065 cm} \end{equation} with $V_{k \neq j}$ being whatever verton. Therefore, we obtain,
\begin{equation}\label{gcr}
\hspace{.11111 cm}  \mathcal{C}_j^{I}\;=\; \prod_{m=1}^{M}\;
\prod_{y_{(m)}}^{{\rm all\;choices}}\;\big[\;\big(1 - \breve{n}_{y_{(m)j}}\big)\;
+\; \breve{r}^\dagger_{Iy_{(m)}}\;\breve{r}_{y_{(m)}j}\;\big]
\end{equation} 
with $R_{Iy_{(m)}}$ being the unique relaton whose indices are the union of $y_{(m)}$ with all the vertonic indices of $F_I$. 
\\ Indeed, one can now directly see that upon the insertion of the \emph{`all-species cascade operator'} (\ref{gcr}), $v_j \mathcal{C}_j^I \breve{f}_I^\dagger$ becomes a hypergraphically-acceptable conversion operator, as its satisfies the constraints (\ref{8abc}). We highlight that, the inverse conversion $F_I \to V_j$ is conducted by its Hermitian-conjugate cascade operator, that is, by $\mathcal{C}_I^j = (\mathcal{C}_j^I)^\dagger$. \emph{The result} (\ref{gcr}) \emph{completes the axiomatic construction of the first-order cascade operators of} $\mathcal{M}$.\\\\
Let us now confirm that indeed principle five has been correctly realized in $\mathcal{M}$. By this principle, the complete set of the second-quantized hypergraphical isomorphisms, $S_M^{(\mathcal{M})}$, must be realized in the theory as specified both in the statement of principle five and in its explanatory note. \emph{First}, the quantum-hypergraphical isomorphisms must be identified with some composite operators which are constructed from a proper subset of the multi-qubit-conversion operators of $\mathcal{M}$, that is, \emph{from a specific class of the Hamiltonian operators} $\mathcal{O}_{\upkappa}^{H^{(\mathcal{M})}_M}$ \emph{as introduced in} (\ref{equa8}). \emph{Second}, every one of these operators must complete a corresponding many-body conversion in $\mathcal{M}$ which is \emph{purely vertonic}. Being mathematically represented, the purely-vertonic conversions do realize all the coordinate permutations on $\mathcal{V}_M^{(\mathcal{M})}$, in a one-to-one manner. Being so, to form the generators $\Gamma_j^i$ (\ref{11vcp}), they must be completed by the induced relatonic adjustments. Looking into $\mathcal{C}_j^i$ in (\ref{sco}), one confirms that, 
\begin{equation}\label{hgi}\begin{split} & \mathcal{G}_j^i \;=\; \mathcal{C}_j^i\;\;\;,\;\;\forall\;\; (j , i)\\ & \Gamma_j^i \;=\;  1 + n_j (1 - n_i) (v_j \mathcal{C}_j^i v^{\dagger}_i - 1) \end{split} \end{equation} Being generated by the operators of the first-order quantum-hypergraphical isomorphisms given in (\ref{hgi}), the `order-$m^{\in \mathbb{N}^{\leq M}}$' elements of $\mathcal{S}_M^{(\mathcal{M})}$ are now constructed, \begin{equation}\label{hohgi}\begin{split} & \Gamma_{j_1 \cdot \cdot \cdot j_m}^{i_1 \cdot \cdot \cdot i_m}\;\equiv\; \Gamma_{j_1}^{i_1}\; \cdot \cdot \cdot\; \Gamma_{j_m}^{i_m}\;=\\ & \hspace{1.26 cm} =\; \prod_{s}^{1 \leq s \leq m}\;[\;1 + n_{j_s}(1 - n_{i_s}) (v_{j_s} \mathcal{C}_{j_s}^{i_s} v^\dagger_{i_s} - 1)\;]\;\\ &\Gamma_{j_1 \cdot \cdot \cdot j_m}^{i_1 \cdot \cdot \cdot i_m}\ket{\Psi}^{\in\;\mathcal{H}_M^{(\mathcal{M})}} =\;\ket{\Psi^\prime}^{\in\;\mathcal{H}_M^{(\mathcal{M})}} \cong\;\ket{\Psi}^{\in\;\mathcal{H}_M^{(\mathcal{M})}}\;\;\;,\; \forall\;\ket{\Psi}^{\in\;\mathcal{H}_M^{(\mathcal{M})}}  \end{split}\end{equation} 
\emph{But, we highlight a very important characteristic of} $\mathcal{M}$. Primarily, the cascade operators $\mathcal{C}_J^I$ are introduced in the theory to secure the hypergraphical well-definedness of HoloQuantum HyperNetworks during their entire time evolutions. Being so, setting aside all the purely-vertonic ones, \emph{the `verton-relaton-conversion cascade operators'}, namely the pairs $(\mathcal{C}_j^{i_{(m)}}, \mathcal{C}^j_{i_{(m)}} )$ in (\ref{gcr}), \emph{do not support any quantum-hypergraphical isometries in} $\mathcal{M}$. That is, their purely-relatonic cascades which complete the verton-relaton conversions $V_j \leftrightarrows R_{i_{(m)}}$, \emph{do transmute the HoloQuantum HyperNetworks `non-isomorphically'}. Let us then state one significant point. \emph{The immensely-larger subset of the many-body interactions in} $\mathcal{M}$ \emph{are all the non-isomorphic  `transmutational interactions'}.\\ 
\emph{Next, we advance to the realization of principle six}. This principle defines the quantum-mechanical redundancies of HoloQuantum HyperNetwork Theory. Re-highlighting a point of distinction, we must remind that principle five is sourced by the mathematical character of $\mathcal{M}$. On the other hand, principle six originates from the physical character of $\mathcal{M}$, being a theory of quantum mechanics. This is simply the statement that the global phases of the basis-states of $\mathcal{H}_M^{(\mathcal{M})}$ must be unobservables. Because all the multi-qubit states of $\mathcal{M}$ are created as (\ref{svs},\ref{aub},\ref{equa7}) identify, the following complete set of the global $U(1)$ transformations on the hypergraph-state-qubits $F_I$s, \begin{equation}\label{u1}
\breve{f}_I \to \exp(-i \phi)\; \breve{f}_I\;\;\;;\;\;\;\breve{f}_I^\dagger \to \exp(+ i \phi)\; \breve{f}_I^\dagger\;\;\;,\;\forall\;\phi \in \mathbb{R}\;,\;\forall\; I
\end{equation} 
which develop those global phases, must be symmetry transformations. 
All the 
observables of the 
theory must be the singlets of the global $U(1)$ transformations (\ref{u1}) which assign charges $(-1,+1)$ to the operator-doublets $(\breve{f}_I,\breve{f}_I^\dagger)$. Being so, every Hamiltonian operator $\mathcal{O}^{(H_M^{(\mathcal{M})})}$ must be composed from the products of equal numbers of the $\breve{f}_J$s and the $\breve{f}_I^\dagger$s, to be $U(1)$-chargeless.
Imposing this condition on the solutions of the quantum constraints (\ref{8abc}) obtained so far, we conclude that all the following operators are among the microscopic interactions of $\mathcal{M}$,
\begin{equation}\begin{split}
\label{aloha}
& \{\; \mathcal{O}^{(H_M^{(\mathcal{M})})}\;\}\;=\; \{\; \mathcal{O}^{(H_M^{(\mathcal{M})})}[\;\{\breve{f}_J\;;\;\{\breve{f}^\dagger_I\}\;]\;\}\; \supset\\ & \supset\;\{\; \big(\breve{r}_{j_{1_{(s_1)}}} ...\; \breve{r}_{j_{m_{(s_m)}}}\big)\big(\breve{r}^\dagger_{i_{m_{(r_m)}}} ...\; \breve{r}^\dagger_{i_{1_{(r_1)}}}\big)\;
;\; v_j \mathcal{C}_j^I \breve{f}_I^\dagger\;;\;\;\; h.c\;\}
\end{split}\end{equation}
\emph{Result} (\ref{aloha}) \emph{does summarize the fundamental rule of} $\mathcal{M}$. \emph{By this result, the `first-order' Hamiltonian operators of HoloQuantum HyperNetwork Theory are so concluded},
\begin{equation}\begin{split}
\label{aloha1}
& \{\; \mathcal{O}^{(H_M^{(\mathcal{M})})}_{\;({\rm first\;order})}\;\}\; =\;\{\; \mathcal{O}^{(H) I}_{\;(1) J}\;\equiv\;\breve{f}_J\;\mathcal{C}_J^I \; \breve{f}_I^\dagger \;\;\;,\;\;\;\forall\; J \;,\;\forall\; I\; \}
\end{split}\end{equation}
Let us now summarize as follows the cascade operators $\mathcal{C}_J^I$ of the first-order Hamiltonian operators (\ref{aloha1}), 
\begin{equation}
\begin{split}
\label{aloha3}
&\mathcal{C}_J^I \in \{\;\mathcal{C}_j^I\;,\;\mathcal{C}_{\breve{r}_{j_{(n)}}}^{\breve{r}_{i_{(m)}}}\;\}\hspace{5.33 cm}\\
&\mathcal{C}_{\breve{r}_{j_{(n)}}}^{\breve{r}_{i_{(m)}}} = 1\;\;\;;\;\; C_j^j = 1 \hspace{6.92 cm}
\\
&\mathcal{C}_j^{I \neq j} \; = \; \prod_{m=1}^{M}\;
\prod_{w_{(m)}}^{{\rm all\;choices}}\;\big(\; 1\; -\; n_{w_{(m)j}}\;
+\; \breve{r}^\dagger_{Iw_{(m)}} \breve{r}_{w_{(m)}j}\; \big)
\end{split}\end{equation}
By the findings (\ref{aloha1},
\ref{aloha3}), we such obtain the \emph{first-order Hamiltonian of  
HoloQuantum HyperNetwork Theory},
\begin{equation}
\label{h11}
H^{(\mathcal{M})(1-1)}_{M} = \;\sum_{I,J}\; \lambda_J^I \;\mathcal{O}^{(H) I}_{\;(1) J}\; =\;
\sum_{I,J}\; \lambda_J^I\;\; \breve{f}_J \;\mathcal{C}_J^I\; \breve{f}_I^\dagger \;
\end{equation}
where for the evolution to be unitary, we demand, 
\begin{equation}
\lambda_I^J = \bar{\lambda}_J^I
\end{equation}
\emph{Now, we must realize principle eight}. Being \emph{one among the intrinsically-Wheelerian principles of HoloQuantum HyperNetwork Theory}, it does also play a central role in fulfilling the intention of the theory $\mathcal{M}$ to become the fundamental complete quantum many body theory of all quantum natures. Principle eight is the statement that the couplings of all the $\mathcal{O}^{(H_M^{(\mathcal{M})})}_\upkappa$ operators inside the total fundamental Hamiltonian of the theory (\ref{equa8}) must be maximally random.  Applying this to the first-order Hamiltonian of $\mathcal{M}$, namely to the Hamiltonian (\ref{h11}), all of the independent couplings $\lambda_{J}^{I}$ must be random and uncorrelated, defined with their \emph{maximally-free Gaussian distributions},
\begin{equation}\label{mfgrp1}
\mathcal{P}(\lambda_J^I\;|\;\mathring{\lambda}_J^I\;,\hat{\lambda}_J^I) \sim \exp\big(-\frac{
	(\lambda_J^I -\mathring{\lambda}_J^I) ^2}{\hat{\lambda}_J^I}\big)
\end{equation}
The characteristic parameters $(\mathring{\lambda}_J^I, \hat{\lambda}_J^I)$ in (\ref{mfgrp1}), by which the statistical mean values and mean-squared values of $\lambda_J^I$s are respectively determined, take \emph{arbitrary values}.
\\\\ We now classify all the microscopic interactions of the first-order Hamiltonian 
(\ref{h11}), taking care of all the index degeneracies which are contained in the defining summation of $H_M^{(\mathcal{M})}$ in (\ref{h11}). Indeed, for every hypergraph-state-qubit $\breve{F}_I$, there comes an independent chemical-potential $\mu_I \equiv \lambda_I^I$ which is Gaussian random. So, $H^{(\mathcal{M})(1-1)}_{M}$ in (\ref{h11}) takes the unfolded form, 
\;
\begin{equation}
\label{insideh11} 
H^{(\mathcal{M})(1-1)}_{M} \;= \;\sum_I^{{\rm all}}\;\mu_I\;n_I \;+
\hspace{.00155 cm}\;\;\sum_{I \neq J}^{{\rm all}}\; \lambda_J^I\;\breve{f}_J \;\mathcal{C}_J^I\; \breve{f}_I^\dagger  \hspace{1.99 cm} \hspace{.33 cm} \end{equation} 
Being microscopically discerned in terms of all the vertonic-and-relatonic interactions, the Hamiltonian (\ref{insideh11}) can be further microscopically re-expressed as follows,
\begin{equation}\label{minsideh11}\begin{split}
& H^{(\mathcal{M})(1-1)}_{M} \; =\;\sum_{i=1}^{M}\;\mu_i\;n_i \;+\;
\sum_{m = 1}^{M}\;
\sum_{i_{(m)}}^{\abs{\{i_{(m)}\}} = 
{M\choose m}}\;\mu_{i_{(m)}} n_{i_{(m)}}\;+  \hspace{.33 cm}
\\
& \hspace{1.55 cm}\; + \sum_{m, n = 1}^{M}\;\sum_{j_{(s)},i_{(m)}}^{i_{(m)} \neq j_{(s)}}\;\lambda_{j_{(s)}}^{i_{(m)}}\; \breve{r}_{j_{(s)}}  \breve{r}_{i_{(m)}}^\dagger \;+\;h.c\;+ \hspace{.11 cm}
 \\
& \hspace{1.33 cm} \;\;\; +\;\sum_{m = 1}^{M}\;\sum_{i_{(m)}}\;\sum_j\; \lambda_j^{i_{(m)}}\; v_{j} \mathcal{C}_j^{i_{(m)}} \breve{r}_{i_{(m)}}^\dagger \;+\;h.c\;+\hspace{.55 cm}
\\
& \hspace{1.33 cm} \;\; +\; \sum_{j,i}^{i \neq j}\; \lambda_j^i\; v_{j} \mathcal{C}_j^i v_{i}^\dagger \;+\; h.c \hspace{1.99 cm}\hspace{.88cm} 
\end{split}
\end{equation}
\emph{Results} (\ref{h11},\ref{insideh11},\ref{minsideh11}) \emph{conclude the mission of section three}. By invoking principle nine, and directly based upon the fundamental rule, we systematically develop the total unitary evolution of the `flavorless theory' in section four. In \emph{section five}, we complete this constructive procedure, by finally presenting \emph{`the kinematically-and-dynamically Perfected HoloQuantum HyperNetwork Theory'}.
\section{HoloQuantum HyperNetwork Theory: Complete Interactions, The Total Hamiltonian, The Compact Model} The mission of this section is \emph{to complete the unitary quantum dynamics of HoloQuantum HyperNetworks} by formulating all the higher-order many-body interactions of $\mathcal{M}$, and therefore, determining its total Hamiltonian. The road map to systematically get there must have already become clear to a great extent by the results of section three which formulated all the core microscopic interactions of the theory, and concluded its first-order Hamiltonian. But because to accomplish this aim, a hierarchical family of the higher-order cascade operators are to be correctly identified, it will be instructive for us to begin the procedure with carefully constructing the second-order microscopic interactions and  therefore the second-order Hamiltonian 
of the theory.\\
\\ Based on the very same principles, the higher-order many-body interactions of HoloQuantum HyperNetwork Theory must be all constructed by the \emph{`consistent prolifications' of its core operations} as being concluded by its fundamental rule. Being so, to formulate the second-order interactions, we should correctly realize the `two-to-two' conversions between the arbitrarily-chosen `two-sets' $\{\breve{F}_{J_1},\breve{F}_{J_2}\} \leftrightarrow \{\breve{F}_{I_1},\breve{F}_{I_2}\}$. Beginning with the simplest ansatz for these interactions, and albeit up to the redundant permutations in the down-indices and in the up-indices, we now examine the following operators,
\begin{equation}\label{sob}
\hspace{1.09 cm} \mathcal{O}^{I_1 I_2}_{J_1 J_2} \;\equiv\; (\breve{f}_{J_1} \breve{f}_{J_2}) (\breve{f}_{I_2}^\dagger \breve{f}_{I_1}^\dagger)
\end{equation} 
If the interaction operators (\ref{sob}) are purely relatonic, namely converting any two relatons to any two relatons, then by satisfying in this simplest form all the dynamical constraints (\ref{8abc}), they are Hamiltonian operators as such. But if any of these four qubits is vertonic, then the ansatz (\ref{sob}) generically violates the dynamical hypergraphical well-definedness of HoloQuantum HyperNetworks for the same reason explained in section three in the case of the core interactions. Being so, by identifying the needful first-order cascade operators for $\{\breve{F}_{J_1},\breve{F}_{J_2}\} \leftrightarrow \{\breve{F}_{I_1},\breve{F}_{I_2}\}$ and combining them sequentially, we enlarge the bare operators (\ref{sob}) by the corresponding assemblies of their cascade conversions. This direct simple procedure leads us to both of the following Hamiltonian operators, 
\begin{equation}\label{sobsa}\begin{split}
& \mathcal{O}^{(H)(I_1,I_2)}_{\;(2)(J_1,J_2)} \equiv \breve{f}_{J_1} \breve{f}_{J_2} (\mathcal{C}_{J_1}^{I_1} \mathcal{C}_{J_2}^{I_2}) \breve{f}_{I_2}^\dagger \breve{f}_{I_1}^\dagger \equiv \breve{f}_{J_1} \breve{f}_{J_2} \;\mathcal{C}_{J_1 J_2}^{I_1 I_2}\; \breve{f}_{I_2}^\dagger \breve{f}_{I_1}^\dagger\\
& \mathcal{O}^{(H)(I_2,I_1)}_{\;(2)(J_1,J_2)} \equiv \breve{f}_{J_1} \breve{f}_{J_2} (\mathcal{C}_{J_1}^{I_2} \mathcal{C}_{J_2}^{I_1}) \breve{f}_{I_2}^\dagger \breve{f}_{I_1}^\dagger \equiv \breve{f}_{J_1} \breve{f}_{J_2}\; \mathcal{C}_{J_1 J_2}^{I_2 I_1}\; \breve{f}_{I_2}^\dagger \breve{f}_{I_1}^\dagger
\end{split}\end{equation}
Given the `descendant cascade operators' $\mathcal{C}^{I I^\prime}_{J J^\prime}$ defined in (\ref{sobsa}), 
now the operators $\mathcal{O}^{(H)(I,I^\prime)}_{\;(2)(J,J^\prime)}$
satisfy the constraints (\ref{8abc}). \emph{Result (\ref{sobsa}) concludes all the second-order cascade operators and all the second-order interactions of} $\mathcal{M}$ .\\ 
Let us highlight a point here. For every arbitrarily-chosen conversion $\{\breve{F}_{J_1},\breve{F}_{J_2}\} \leftrightarrow \{\breve{F}_{I_1},\breve{F}_{I_2}\}$, there is a freedom in enlarging the operator (\ref{sob}) according to either of the two distinguished channels of the one-to-one conversions which build up this process. This freedom is mirrored in the two independent second-order operators that (\ref{sobsa}) identifies. Therefore, these two independent operators must contribute to the Hamiltonian of the theory with two independent couplings, otherwise the resulted total dynamics of the theory would be restricted unnecessarily. This must be clear because the dimension of the complete space of the second-order interaction operators is counted by the basis-operators which, modulo all the redundant permutations, can be so decomposed in terms of their direct ancestors, 
\begin{equation}\label{sos} 
\begin{split}
&
\{\; \mathcal{O}^{(H) I_1}_{J_1}\; \mathcal{O}^{(H) I_2}_{J_2} \;=\;( \breve{f}_{J_1}\;\mathcal{C}_{J_1}^{I_1} \; \breve{f}_{I_1}^\dagger)\; ( \breve{f}_{J_2}\;\mathcal{C}_{J_2}^{I_2} \; \breve{f}_{I_2}^\dagger)\;;\;\forall \vec{J}\; ,\; \forall \vec{I}\; \;\} \hspace{.45 cm} \hspace{1.99 cm} 
\end{split}\end{equation}
Therefore, by the counting of (\ref{sos}), the Hamiltonian which generates the complete dynamics of the theory should receive an independent coupling for each one of the two operators identified in (\ref{sobsa}). By the result (\ref{sobsa}), and again by the realizition of principle eight, we obtain as follows \emph{the second-order Hamiltonian of HoloQuantum HyperNetwork Theory}, \begin{equation}\label{h22}
\begin{split}
&H^{(\mathcal{M})(2-2)}_{M}
=\;\sum_{J_1,J_2}\;\sum_{I_1,I_2}\;\; \lambda_{J_1J_2}^{I_1I_2}\;\; \breve{f}_{J_1}\breve{f}_{J_2}\;  \mathcal{C}_{J_1J_2}^{I_1I_2}\; \breve{f}_{I_2}^\dagger \breve{f}_{I_1}^\dagger
\end{split}
\end{equation}
with three structural patterns on the $\lambda$-couplings. \emph{First}, to ensure the unitarity of the evolution,
\begin{equation}\label{rlu}
\lambda_{I_1I_2}^{J_1J_2} = 
\bar{\lambda}_{J_1J_2}^{I_1I_2}
\end{equation}
\emph{Second}, to undo the permutational redundancies, 
\begin{equation}
\lambda_{J_2J_1}^{I_1I_2} = 
- \lambda_{J_1J_1}^{I_1I_2}\;\;\;;\;\;\; \lambda_{J_1J_2}^{I_2I_1} = - \lambda_{J_2J_1}^{I_1I_2}
\end{equation}
\emph{Third}, absolutely all the independent couplings $\lambda_{J_1J_2}^{I_1I_2}$ are the maximally-free Gaussian random parameters, to be taken from the same distributions as in (\ref{mfgrp1}).\\ 
\\ By opening-up the hypergraph-state-qubit content of (\ref{sobsa}), we obtain the microscopically-distinguished second-order many-body interactions of HoloQuantum HyperNetwork Theory as follows, \begin{equation} 
\begin{split}\label{ufl}
&\mathcal{O}^{(H)}_{(J_1,J_2),(I_1,I_2)}\;\in \; \{\;(\breve{r}_{j_{1_{(s_1)}}} \breve{r}_{j_{2_{(s_2)}}}) (\breve{r}^\dagger_{i_{2_{(m_2)}}} \breve{r}^\dagger_{i_{1_{(m_1)}}}) \;;\;h.c\;;\;
\\
&\;(v_{j_1} \breve{r}_{j_{2_{(s_2)}}})\; \mathcal{C}_{j_1 j_{2_{(s_2)}}}^{i_{1_{(m_1)}}i_{2_{(m_2)}}}\; (\breve{r}^\dagger_{i_{2_{(m_2)}}} \breve{r}^\dagger_{i_{1_{(m_1)}}})\;;\;h.c \;;
\\
&\;(v_{j_1} v_{j_2})\; \mathcal{C}_{j_1 j_2}^{i_{1_{(m_1)}}i_{2_{(m_2)}}}\; (\breve{r}^\dagger_{i_{2_{(m_2)}}} \breve{r}^\dagger_{i_{1_{(m_1)}}})\;;\;h.c \;;
\\
&\;(v_{j_1} v_{j_2})\; \mathcal{C}_{j_1 j_2}^{ i_{1_{(m_1)}} i_2}\; (v^\dagger_{i_2} \breve{r}^\dagger_{i_{1_{(m_1)}}})\;;\;h.c \;;
\\
&\;(v_{j_1} v_{j_2})\; \mathcal{C}_{j_1 j_2}^{i_1 i_2}\; (v^\dagger_{i_2} v^\dagger_{i_1})\;;\;h.c\;\}
\end{split} \end{equation} 
Having manifested (\ref{ufl}), let us further expose all the microscopically-distinct terms in (\ref{h22}). By taking care of all the qubit degeneracies in the summations of (\ref{h22}), and by appropriately renaming some of the couplings,  the microscopically-detailed form of (\ref{h22}) is so obtained,   
\begin{equation}\label{h222} 
\begin{split}
& H^{(\mathcal{M})(2-2)}_{M} =  \sum_{I_1,I_2}\; \lambda_{I_1I_2}\; \breve{n}_{I_1}\breve{n}_{I_2}\;\; + \hspace{3.99cm}
\\
&\hspace{1.55 cm}+\;\sum_K\; \sum_{J,I}^{J \neq I}\; \lambda_{J,I}^{K}\;\breve{n}_{K}\;(\breve{f}_{J}\;\mathcal{C}_{J}^{I}\; \breve{f}_{I}^\dagger)\; +\hspace{1.55cm}
\\
&\hspace{1.55 cm} + \sum_{J_1,J_2,I_2,I_1}^{\{J_s\}\cap\{I_r\}=0} \; \lambda_{J_1J_2}^{I_1,I_2}\; \breve{f}_{J_1}\breve{f}_{J_2} \;\mathcal{C}_{J_1,J_2}^{I_1,I_2}\; \breve{f}_{I_2}^\dagger \breve{f}_{I_1}^\dagger \;
\end{split} 
\end{equation} 
By the first class of interaction terms in (\ref{h222}), the number operators of all possible pairs of `HyperNetwork qubits' are coupled Gaussian-randomly and `index-globally'. By the interactions of the second term of (\ref{h222}) the one-to-one conversions of any two distinct qubits, coupled with the number operator of an arbitrary qubit, are turned on Gaussian-randomly. Finally, the interactions of the third class of Hamiltonian terms in (\ref{h222}) consist of all the possible two-to-two conversions of four distinct qubits, activated by the global and Gaussian-random couplings.\\ 
\\ Next, the higher-order Hamiltonians of the theory must be all built up in the exact same way $H_M^{(\mathcal{M})(1-1)}$ and $H_M^{(\mathcal{M})(2-2)}$ have been constructed. Because by now all the steps are crystal clear, we summarize the final result. \emph{The `order-$m$' Hamiltonian of HoloQuantum HyperNetwork Theory, $H^{(\mathcal{M})(m-m)}_{M}$, is so concluded},
\begin{equation}\label{h245} H^{(\mathcal{M})(m-m)}_{M} = \sum_{J_1 ... J_m} \sum_{I_1 ... I_m} \lambda_{J_1 \cdot \cdot \cdot J_m}^{I_1\cdot \cdot \cdot  I_m} (\prod_{s = 1}^{m} \breve{f}_{J_{s}})\;\mathcal{C}_{J_1 \cdot \cdot \cdot J_m}^{I_1\cdot\cdot\cdot I_m} (\prod_{r=1}^{m} \breve{f}_{I_r}^\dagger)
\end{equation}
in which the descendant \emph{`order-m cascade  operators'}, $\mathcal{C}^{\vec{I}}_{\vec{J}}\; \equiv\; 
\mathcal{C}^{I_1 ... I_m}_{J_1 ... J_m}\;$, being required to have the many-body interaction operators of $\mathcal{M}$ satisfying (\ref{8abc}), are identified by means of their  ancestors 
$C_J^I$ given in (\ref{aloha3}) 
simply as, 
\begin{equation}\label{mco}  \mathcal{C}_{J_1 \cdot \cdot \cdot J_m}^{I_1 \cdot \cdot \cdot I_m}\;\equiv\;  \prod_{s}^{1 \leq s \leq m} \mathcal{C}_{J_{s}}^{I_{s}} \end{equation}
The couplings $\lambda_{J_1\cdot \cdot \cdot J_m}^{I_1\cdot \cdot \cdot  I_m}$ in (\ref{h245}) are antisymmetric both in their down-indices and in their up-indices, and also for unitarity satisfy the $m$-body version of (\ref{rlu}). Namely,
\begin{equation}\label{h345} 
\lambda_{J_1 \cdot \cdot \cdot J_m}^{I_1 \cdot \cdot \cdot I_m} = \lambda_{[J_1 \cdot \cdot \cdot J_m]}^{I_1 \cdot \cdot \cdot I_m} = \lambda_{J_1 \cdot \cdot \cdot J_m}^{[I_1 \cdot \cdot \cdot I_m]}\;;\; \lambda_{I_1 \cdot \cdot \cdot I_m}^{J_1 \cdot \cdot \cdot J_m} \;=\;  \bar{\lambda}_{J_1 \cdot \cdot \cdot J_m}^{I_1 \cdot \cdot \cdot I_m}\end{equation}
Besides (\ref{h345}), all the independent couplings $\lambda_{J_1\cdot \cdot \cdot J_m}^{I_1\cdot \cdot \cdot  I_m}$, 
\emph{at the fundamental 
level 
of 
the 
theory}, 
are uncorrelatedly maximally-free Gaussian-random. Namely,
\begin{equation}\label{h445}
\mathcal{P}(\lambda_{\vec{J}}^{\vec{I}}\; |\; \mathring{\lambda}_{\vec{J}}^{\vec{I}}\; , \hat{\lambda}_{\vec{J}}^{\vec{I}}) \sim 
\exp\big(
-\frac{ \big(\lambda_{\vec{J}}^{\vec{I}}- \mathring{\lambda}_{\vec{J}}^{\vec{I}}\big)^2 }
{ \hat{\lambda}_{\vec{J}}^{\vec{I}}}
\big)
\end{equation}
Now we are at \emph{the right 
point 
to meet principle nine}.\\
\\ {\bf Principle 
9:}  $\mathcal{M}$, for being the `it-from-qubit' theory of all quantum natures and all possible HoloQuantum Hypergraphs, must be `\emph{covariantly complete}'.\\\\ \emph{Explanatory Note:} Principle nine, by being the statement of \emph{`covariant completeness'}, is a must to secure the fulfillment of the intention of HoloQuantum HyperNetwork Theory. By this principle, \emph{being realized on the `multiverse face' of} $\mathcal{M}$, every quantum (or even classical) many body system of the arbitrarily-chosen objects and their arbitrarily-chosen relations, which is in-principle realizable, must be both completely and form-invariantly (=  covariantly) formulable by this very theory. That is, the complete time-dependent information  of all the observables and all the states of all of these many body systems must be \emph{either} extractable from the complete theory as its covariant contextual model-specifications or consistent solutions, \emph{or} be contained in its total phase diagram as the covariant effective-or-emergent (sub)theories. \emph{Realized on its mathematical `hypergraphical face'}, this principle states that, every dynamical hypergraphical (and so also graphical) network which is both structurally and functionally consistent must be 
covariantly formulable from within the this complete theory, either directly as extractions or solutions, or as its phase-specific effective emergences.\\ 
\\ For principle nine to come true, given the absolute generality of our principles and all the formulation, there is \emph{only one condition} we still need to impose. By the Wilsonian renormalization group flow, the landscape of the low-energy fixed-points of the theory must be, both physics-wise and network-wise, covariantly complete. This `low-energy completeness' of the theory requires that, all the relevant-or-marginal multi-qubits-conversion operators which are acceptable by principles four, five, six and seven, must contribute, in the way stated by principle eight, to the total Hamiltonian (\ref{equa8}). So, \emph{the total Hamiltonian of} $\mathcal{M}$, $H_M^{(\mathcal{M})}$, \emph{must be form-invariant under the whole Wilsonian renormalization group flow}.\\ \\ Let us now realize principle nine by which the complete microscopic interactions and the total Hamiltonian of HoloQuantum HyperNetwork Theory
$H_M^{(\mathcal{M})}$ must be obtained. By principle two, $\mathcal{M}$ lives in $(0+1)$ dimensions. Exactly in these dimensions, all the $M^\star$ hypergraph-state `alphabet operators' $(\breve{f}_I, \breve{f}_I^\dagger)$ have canonical dimension zero, by formally being fermions. As such, all the acceptable convertors, no matter how `large' they are as fermionically-made composite bosonic operators, are relevant, in fact, are `equally relevant'. So, $H^{(\mathcal{M})}_M$ must be formed hierarchically by receiving all $\mathcal{O}^{(H)}_{(m-m)}$s. Namely, to realize principle nine, one must sum up $H^{(\mathcal{M})(m-m)}_{M}$, $\forall\;m_{\geq 1}^{\leq M^\star}$, to obtain $H_M^{(\mathcal{M})}$. 
\\ Therefore, \emph{the realization of principle nine is so fulfilled}, 
\begin{equation}\label{h111be11b} H^{(\mathcal{M})}_{M} = \sum_{m=1}^{m=M^\star} H_M^{(\mathcal{M})(m-m)} \end{equation}
Now, having realized all the nine principles, and given the $m$-body results (\ref{h245},\ref{h345},\ref{h445}), we conclude as follows. \emph{The complete unitary evolution operator of HoloQuantum HyperNetworks}, $U_M^{(\mathcal{M})}(t)$, \emph{is generated in accordance with} (\ref{equa8}), \emph{by the following total Hamiltonian},
\begin{equation}\label{h111be11}\begin{split}
& H^{(\mathcal{M})}_{M} = \\ & = \sum_{m=1}^{m=M^\star} \{ \sum_{J_1 ... J_m} \sum_{I_1 ... I_m} %
		\lambda_{J_1\cdot \cdot \cdot J_m}^{I_1\cdot \cdot \cdot  I_m}\; (\prod_{s = 1}^{m} \breve{f}_{J_{s}})\;\
(\prod_{\ell =1}^{m} \mathcal{C}_{J_\ell}^{I_\ell})\;(\prod_{r=1}^{m} \breve{f}_{I_r}^\dagger)\; \} \\
\end{split} \end{equation}
Let us make explicit \emph{the microscopic content} of (\ref{h111be11}). By discerning all the index degeneracies in the summation, and upon renaming some of the the couplings, the total Hamiltonian (\ref{h111be11}) can be re-expressed as follows, 
\begin{equation}\label{ube111}\begin{split}
& \hspace{.0199 cm} H_{M}^{(\mathcal{M})}\; = \;\sum_{m=1}^{m=M^\star}\; \{\; \sum_{I_r}^{1\leq r\leq m}\lambda_{I_1 ... I_m}\;\breve{n}_{I_1}\;...\; \breve{n}_{I_m}\;+\hspace{7.99 cm} \\
& \hspace{.0199 cm} \hspace{1.99 cm} \hspace{.33 cm} + \sum_{I_r \neq J_s}^{1\leq s, r \leq m} \lambda_{\vec{J}}^{\vec{I}}\;\;(\prod_{s=1}^{m} \breve{f}_{J_s})\; \mathcal{C}_{\vec{J}}^{\vec{I}}\;(\prod_{r=1}^{m} \breve{f}_{I_r}^\dagger)\;+\;\hspace{3.99 cm}\hspace{1.99 cm}
\\
& \hspace{.0199 cm} \hspace{.92 cm} \hspace{1.33 cm} +\;
\sum_{p=1}^{m}\sum_{K_c}^{ 1\leq c\leq p}\; 
\sum_{I_r \neq J_s}^{1\leq s, r \leq m-p} \lambda_{\vec{J},\vec{I}}^{\vec{K}} \;(\prod_{c=1}^{p} \breve{n}_{K_c})\;\times\hspace{6.55cm}
\\
& \hspace{3.33 cm}\hspace{.0165 cm} \hspace{1 cm} \times\; (\prod_{s=1}^{m-p} \breve{f}_{J_s})\; \mathcal{C}_{\vec{J}}^{\vec{I}}\;(\prod_{r=1}^{m-p} \breve{f}_{I_r}^\dagger)\; \}
\end{split} \end{equation} 
That is, HoloQuantum HyperNetwork Theory realises three microscopically-distinguished classes of abstract $m$-body hypergraph-state-qubit interactions, for every $m_{\geq 1}^{\leq  M^\star}$.
By the first class, (only) the number operators of $m$ hypergraph-state-qubits interact randomly. By the second class, $m$ one-to-one conversions between a group of all-distinguished hypergraph-state-qubits, together with their cascade conversions, are triggered randomly. By the third class, both of the above-mentioned types of interactions are simultaneously merged in arbitrary mixtures and by random strengths.\\
\\ We re-highlight three important points about $H^{(\mathcal{M})}_{M}$. \emph{First}, in $\mathcal{M}$, interactions between the networkical qubits are `all-species-inclusive'. Namely, the multi-converting qubits can be all vertons, can be a number of relatons of the same-or-different degrees, or can be any arbitrary mixtures of vertons and relatons.\\ 
\\ \emph{Second}, In $\mathcal{M}$, all the HyperNetwork qubits with all possible indices interact, at the fundamental level. That is, the many-body interactions are index-global.\\ 
\emph{Third}, when one comes to develop application-specific sub-theories or models of $\mathcal{M}$, all the types of necessary \emph{disciplines on the random couplings} must be imposed, by simply choosing so, by imposing extra symmetries, or by the emergences. For example, in many sub-theories, solutions or phases of the theory, the `unfrozen' $\hat{\lambda}_{\vec{J}}^{\vec{I}}$ are reduced to only specific subsets of the HyperNetwork qubit spices or their indices. Moreover, the parameters of the Gaussian distributions (\ref{h445}) of the unfrozen couplings must be chosen, or be effectively developed, such that the characteristic statistical-average-measures of those couplings obey the necessary conditions or constraints, given the context. As a physically-significant example, in the models, solutions or phases of the theory in which \emph{geometric locality} is either a built-in assumption or an emergent feature, usually a set of \emph{locality constraints} on the distribution-fixing parameters of those random couplings should ensure that their associated statistical measures do properly depond on the metric-induced distances in 
the corresponding 
geometry.\\
\\ Before moving to the next section, we must now feature \emph{a remarkable sub-theory of the complete theory} (\ref{h111be11}). We will name it the \emph{`Compactified HoloQuantum HyperNetwork Theory'}, and will likewise denote it by $\mathcal{M}_{\bigcirc}$. Both structurally and qualitatively, $\mathcal{M}_{\bigcirc}$  must be regarded as \emph{the first `child' of HoloQuantum HyperNetwork Theory} $\mathcal{M}$. $\mathcal{M}_{\bigcirc}$ as a sub-theory of $\mathcal{M}$, is 
remarkable by being both \emph{`the maximal one'} and \emph{`the minimal one'}. Being qualified as a maximal sub-theory, its definition features the following characteristics. \emph{First}, the entirety of the hypergraph-state-qubits $\breve{F}_I$ are included in its quantum statics, so that its total Hilbert space is identical to (\ref{equa7}). \emph{Second}, the complete abstract many-body interactions of HoloQuantum HyperNetwork Theory are kept activated in its dynamics. That is,
\begin{equation}\label{hbc}\begin{split}
&\hspace{2.09 cm} \mathcal{H}_{M}^{(\mathcal{M}_{\bigcirc})} \;\equiv\; \mathcal{H}_{M}^{(\mathcal{M})}\\ & 
\{\;\mathcal{O}^{(H_M^{(\mathcal{M}_{\bigcirc})})}_{({\rm order}-m)}\;\;;\;\forall\;m_{\geq 1}^{\leq M^\star}\; \}\;\equiv\;\{\;\mathcal{O}^{(H_M^{(\mathcal{M})})}_{({\rm order}-m)} \;\;;\;\forall\;m_{\geq 1}^{\leq M^\star}\; \}
\end{split}\end{equation} 
Being qualified at the same time as a minimal sub-theory, its definition features one more characteristics as follows. The total number of independent couplings in the total Hamiltonian of $\mathcal{M}_{\bigcirc}$ are kept as minimal as possible, with the constraint that (\ref{hbc}) still hold. To realize that, this sub-theory is dynamized by a total Hamiltonian which exponentiates the first-order Hamiltonian (\ref{h11}). Namely, 
\begin{equation}\label{child}\begin{split}
& H^{(\mathcal{M}_{\bigcirc})}_{M}\; \equiv\; \exp \big(H^{(\mathcal{M})(1-1)}_M \big) 
 \hspace{.099 cm} =\; \exp\big(\sum_{I,J} \lambda_J^I\; \breve{f}_J \;\mathcal{C}_J^I\; \breve{f}_I^\dagger\big)
\end{split} \end{equation}
The above Hamiltonian of $\mathcal{M}_{\bigcirc}$ which (up to the addition of an irrelevant constant term) defines a sub-theory of (\ref{h111be11}), realizes a campactification of the moduli space of the Gaussian random couplings of the complete theory $\mathcal{M}$ 
to that subspace 
which triggers 
(\ref{h11}).
\section{HoloQuantum HyperNetwork Theory:  
The Maximally-Flavored Formulation, 
`The Perfected Theory'} \emph{Now, HoloQuantum HyperNetwork Theory $\mathcal{M}$ must become `perfected', based on the very nine principles}. \emph{On one hand}, this perfection amounts to arriving at the \emph{`absolute' mathematical generality
of} $\mathcal{M}$, by formulating the HoloQuantum hypergraphs-and-HyperNetworks which are maximally flavored, both relatonically and vertonically. By this ultimate mathematical enlargement of $\mathcal{M}$, most particularly, all possible HoloQuantum HyperNetworks (and so hypergraphs) whose vertons (and so quantum vertices) and relatons (and so quantum hyperlinks) are in the most general form \emph{`weighted'} and \emph{`oriented'} are formulated. Moreover, this goal will be accomplished in the very `it-from-qubit' natural way. However we must highlight that the identifications of these flavors are not restricted to the hypergraphical weights or orientations. Most-generally understood, the maximally-falvored relatons and vertons frepresent all possible families of quantum hyperlinks and quantum vertices which, although can take up all the different identification (when being interpreted contextually), are nevertheless the constituting and interacting degrees of freedom of the `Perfected $\mathcal{M}$'. \emph{On the other hand}, this maximally-flavored formulation amounts to the \emph{ultimate realization of principle nine in the physics-face of} $\mathcal{M}$. This is so because the resulted maximally-flavored $\mathcal{M}$ does formulate, as is intended, \emph{`absolutely' all possible quantum many body systems of quantum objects and their quantum relations, namely, all quantum natures}.\\
\\ We begin to construct the total quantum kinematics, namely $\mathcal{H}^{(\mathcal{FM})}_{M}$, and the complete quantum dynamics, namely $H^{(\mathcal{FM})}_M$, of the \emph{`Maximally-Flavored $\mathcal{M}$'}. Modulo defining the flavored cascade operators, the procedure is straghtforward, nevertheless it perfects qualitatively $\mathcal{M}$.\\\\ \emph{First}, we enlarge the Hilbert space $\mathcal{H}_{M}^{(\mathcal{M})}$ of section three, by replacing its relatons with the \emph{`upgraded relatons'} which, for every choice of the base-vertons, are \emph{flavored in the most general form}. Clearly, HyperNetworks can, and in fact do generically, develop the different `classes = flavors' of hyperlinks which represent the distinct types of relations between the same vertex-represented objects. Mathematically, for example, all possible weights or orientations of every single hyperlink can be formulated as certain flavors, as we will soon manifest. Physically, on the other hand, the maximally-flavored relatons are qubits for the `distinct hyperlinks' which represent all the `distinct-type' many-body interactions,  many-body correlations, or the structural-or-functinal many-body compositions, connections, or participatig associations. Being maximally general, \emph{we let every relaton $R_{i_{(m)}}$ be upgraded by its most general set of the `purely-relatonic flavors'} $=\;{\rm Set}^{{R}_{i_{(m)}}}\big({\alpha[{i_{(m)}]}}\big)$, \emph{as} $R_{i_{(m)}}^{\alpha[{i_{(m)}}]}$.\\ 
\emph{Second}, we must likewise enlarge $\mathcal{H}_{M}^{(\mathcal{M})}$ by replacing its flavorless vertons with \emph{the `upgraded vertons' which are most-generally flavored}. By definition, the such-flavored vertons represent (as before,) information-theoretically all possible defining-or-refining states which can be associated to the arbitrarily-chosen objects, either each by each distinctly, or in the arbitrarily-chosen groups. For example, on one hand they can represent the weights or the colors of the vertices, and on the other hand, the spins, the bosonic states of different population numbers, the `generations', and any quantum numbers of particles. 
Being maximally general, therefore, \emph{we let every single verton $V_i$ be now upgraded by its most general set of the `purely vertonic flavors'} $=\;{\rm Set}^{V_{i}}\big(a_{i}\big)$, \emph{as} $V_i^{a_i}$. For consistency, however, these purely-vertonic flavors also induce their  `vertonically-induced flavors' on relatons. This must be clear, because relatons are defined to be the information-theoretically-defining qubits of the quantum hyperlinks which must be uniquely identified by their now-flavored quantum base-vertices. Being so, aside from its purely-relatonic flavors, every relaton msut be additionally carrying a whole sequence of the purely-vertonic flavors which define its base-vertons, to be identified uniquely.  Collecting all the data, we denote \emph{the all-flavores-included $m$-relatons} by $R_{(i^{a_i})_{{}_{(m)}}}^{\alpha[(i^{a_i})_{{}_{(m)}}]}$.\\ \\ \emph{Third}, to have the maximally-flavored theory of $\mathcal{M}$ 
be formulated with notational compactness, assuming that the above index-details are all \emph{remembered}, the maximally-flavored hypergraph-state-qubit operators will be collectively denoted by $\big(\breve{f}_I^{\gamma_{{}_{I}}}, \breve{f}_I^{\dagger \gamma_{{}_{I}}}\big)$, in which $I^{\in  \mathbb{N}^{\leq M^\star}}$ is the all-inclusive counter of the very flavorless qubits. Now, \emph{the total state-space of 
Maximally-Flavored HoloQuantum HyperNetwork Theory}, 
$\mathcal{H}_{M}^{(\mathcal{FM})}$, \emph{is defined by its tensor-product structure and by its basis} $\mathcal{B}_{M}^{(\mathcal{FM})}$, \begin{equation}\label{equa71}\begin{split}
& \mathcal{H}_{M}^{(\mathcal{FM})}\; =\; \mathcal{H}_{(\rm{vertons})}^{(\mathcal{FM})}\; \otimes_{{}_{1 \leq m \leq M}}\; \mathcal{H}_{(m- {\rm relatons})}^{(\mathcal{FM})}\\
& \mathcal{B}_{M}^{(\mathcal{FM})} \;=\;
\big\{\;\prod_{I_s}^{{\rm all\;choices}}\;\;\prod_{\gamma_{{}_{I_s}}}^{{\rm all\;choices}}\;\; \breve{f}_{I_s}^{\dagger 
\gamma_{{}_{I_s}}}\;\ket{0} 
\;\}
\end{split}\end{equation}
The identification (\ref{equa71}) is the quantum-kinematical maximally-flavored realization of principles \emph{one to three}. Reading directly from (\ref{equa71}), the very maximally-flavored formulation of principles \emph{seven, eight and six} must be clear. Specially, the U(1) redundancy transformations of the maximally-flavored qubits are so formulated,
\begin{equation}\label{u71}
\forall\; (I,\gamma_{{}_{I}}):\; \breve{f}_I^{\gamma_{{}_{I}}} \to e^{-i \phi}\; \breve{f}_I^{\gamma_{{}_{I}}}\;\;\;;\;\;\;\breve{f}_I^{\dagger \gamma_{{}_{I}}} \to e^{+ i \phi}\; \breve{f}_I^{\dagger \gamma_{{}_{I}}}
\end{equation}
Next, we move on to formulate systematically \emph{the total unitary dynamics of the perfected theory}. Building on all the results of the previous sections, this mission must be accomplished as straightforwardly as possible. Being so, the major step to obtain $H_M^{(\mathcal{FM})}$ is to realize principles \emph{four and five}
by defining the \emph{`maximally-flavored cascade operators'}, which generalize (\ref{aloha3},\ref{mco}) correctly.
\\ Remembering the notations $\{\breve{F}_I^{\gamma_{{}_{I}}}\} = \{ V_i^{a_i} ; R_{(i^{a_i})_{{}_{(m)}}}^{\alpha[(i^{a_i})_{{}_{(m)}}]} \} $, \emph{the Maximally-Flavored Cascade Operators} $\mathcal{C}_{J^{\gamma_{{}_{J}}}}^{I^{\eta_{{}_{I}}}}$, \emph{and all their order-$m$ descendants}, are so identified correctly,
\begin{equation}\begin{split}
\label{unifyg71}
& \{\; \mathcal{C}_{J^{\gamma_{{}_{J}}}}^{I^{\eta_{{}_{I}}}}\;\} \;\equiv\; \{ \mathcal{C}_{j^{a_j}}^{I^{\eta_{{}_{I}}}}(\upmu_{j^{a_j} }^{I^{\eta_{{}_I}}})\;;\;{\rm All\;the\;purely\;relatonic\;ones} = 1 \}
\hspace{6.92 cm}\hspace{5.33 cm} \\ &\; \hspace{3 cm}  \\
& \mathcal{C}_{j^{a_j} }^{I^{\eta_{{}_I}}} (\upmu_{j^{a_j} }^{I^{\eta_{{}_I}}}) = \\ & = \; \prod_{m=1}^{M}
\prod_{(y^{a_y})_{{}_{(m)}}}^{{\rm all\;choices}} \prod_{\alpha[(y^{a_y})_{{}_{(m)}} j^{a_j}]}^{{\rm all\;choices}} \mathcal{C}_{\alpha[(y^{a_y})_{{}_{(m)}} j^{a_j}]}^{j^{a_j}\; \to\; I^{\eta_{{}_I}}}\big[ \upmu_{\alpha[(y^{a_y})_{{}_{(m)}} j^{a_j}]}^{\beta[I^{\eta_{{}_I}}(y^{a_y})_{{}_{(m)}}]
 } \big] \\ & \\ & \mathcal{C}_{\alpha[(y^{a_y})_{{}_{(m)}} j^{a_j}]}^{j^{a_j}\; \to\; I^{\eta_{{}_I}}}\big[ \upmu_{\alpha[(y^{a_y})_{{}_{(m)}} j^{a_j}]}^{\beta[I^{\eta_{{}_I}}(y^{a_y})_{{}_{(m)}}]
 }\big]\;\;= \\ & \hspace{1.99 cm} \hspace{1.99 cm} \hspace{.69 cm} =\; \big(\;1\; -\; 
\breve{n}_{(y^{a_y})_{{}_{(m)}} j^{a_j}}^{\alpha[(y^{a_y})_{{}_{(m)}} j^{a_j}]}
\;\;+\\
&\hspace{.00115 cm}\hspace{.00126 cm}
 + \sum_{\beta[I^{\eta_I} (y^{a_y})_{{}_{(m)}}]}^{\rm{all\;choices}} \mu_{\alpha[(y^{a_y})_{{}_{(m)}} j^{a_j}]}^{\beta[I^{\eta_{{}_I}}(y^{a_y})_{{}_{(m)}}]
 }\;\;
 \breve{r}_{I^{\eta_{{}_I}}(y^{a_y})_{{}_{(m)}}}^{\dagger \beta[I^{\eta_{{}_I}}(y^{a_y})_{{}_{(m)}}]}
\; \breve{r}_{(y^{a_y})_{{}_{(m)}} j^{a_j}}^{\alpha[(y^{a_y})_{{}_{(m)}} j^{a_j}]} \big)\;\\ & \;\;\;\hspace{1.773 cm}  \\
& \upmu_{\alpha[(y^{a_y})_{{}_{(m)}} j^{a_j}]}^{\beta[I^{\eta_{{}_I}}(y^{a_y})_{{}_{(m)}}]}\;\equiv\;\; \bigcup_{\beta[I^{\eta_{{}_I}}(y^{a_y})_{{}_{(m)}}]}^{{\rm all\;choices}}\; \{\; \mu_{\alpha[(y^{a_y})_{{}_{(m)}} j^{a_j}]}^{\beta[I^{\eta_{{}_I}}(y^{a_y})_{{}_{(m)}}]}\; \} \\ 
& \upmu_{j^{a_j} }^{I^{\eta_{{}_I}}}\; \equiv\;\; \bigcup_{m = 1}^{M}\; \bigcup_{(y^{a_y})_{{}_{(m)}}}^{{\rm all\;choices}}\;  \bigcup_{\alpha[(y^{a_y})_{{}_{(m)}} j^{a_j}]}^{{\rm all\;choices}}\; \upmu_{\alpha[(y^{a_y})_{{}_{(m)}} j^{a_j}]}^{\beta[I^{\eta_{{}_I}}(y^{a_y})_{{}_{(m)}}]}  \\ & 
\\ 
& \mathcal{C}_{J_1^{\gamma_{{}_{{}_{J_1}}}} \;\cdot \cdot \cdot\; J_m^{\gamma_{J_m}}}^{I_1^{\eta_{{}_{I_1}}} \;\cdot \cdot \cdot\; I_m^{\eta_{{}_{I_m}}}} \;=\;\;\mathcal{C}_{J_1^{\gamma_{{}_{J_1}}}}^{I_1^{\eta_{{}_{I_1}}}} \; \cdot \cdot \cdot\; \mathcal{C}_{J_m^{\gamma_{{}_{J_m}}}}^{I_m^{\eta_{{}_{I_m}}}} 
\end{split} 
\end{equation}
Here comes the identification of all the characters which participate in the generalized definitions given by (\ref{unifyg71}). Every index-set $(y^{a_y})_{{}_{(m)}} \equiv\;\{y_1^{a_{y_1}} ... \;y_m^{a_{y_m}}\}$ identifies an arbitrarily-chosen set of $m_{\geq 1}^{\leq M-2}$ `intermediate vertons' $\{V_{y_1}^{a_1} \cdot \cdot \cdot V_{y_m}^{a_m}\}$. Being highlighted, the second product in (\ref{unifyg71}) incorporates the largest spectra of the flavors which must be attributed to the intermediate vertons, for them to be identified uniquely. Given the chosen $(y^{a_y})_{{}_{(m)}}$, every $\breve{R}_{(y^{a_y})_{{}_{(m)}} j^{a_j}}^{\alpha[(y^{a_y})_{{}_{(m)}} j^{a_j}]}$ is the uniquely-identified relaton whose base indices are given by $\{j^{a_j}\} \cup (y^{a_y})_{{}_{(m)}}$, and has the purely-relatonic identity-flavor `$\alpha$'. Given $j^{a_j}(y^{a_y})_{{}_{(m)}}$, the third product in (\ref{unifyg71}) incorporates all the `such-based relatons' which are accommodated in the largest spectra of the purely-relatonic identity-flavors. $\breve{R}_{I^{\eta_{{}_I}}(y^{a_y})_{{}_{(m)}}}^{\beta[I^{\eta_{{}_I}}(y^{a_y})_{{}_{(m)}}]}$ is, likewise, the unique relaton whose base is given by the union of $(y^{a_y})_{{}_{(m)}}$ with all the vertonic indices of $I^{\eta_{{}_{I}}}$, and is specified by the purely-relatonic identification-flavor `$\beta$'. By the intermediate summation in (\ref{unifyg71}), all the relatons with the base $I^{\eta_{{}_I}}(y^{a_y})_{{}_{(m)}}$ in the largest spectra of their purely-relatonic identity-flavors must be integrated.\\ We complete the identification of the characters in (\ref{unifyg71}), by specifying the \emph{`$\mu$-couplings'} which, at the level of the fundamental formulation of the perfected theory, come in the definitions of its maximally-flavored cascade operators $\mathcal{C}_{j^{a_j}}^{I^{\eta_{{}_{I}}}}(\upmu_{j^{a_j} }^{I^{\eta_{{}_I}}})$. Upon realizing principle eight,
for every `initial' purely-relatonic flavor $\alpha[(y^{a_y})_{{}_{(m)}} j^{a_j}]$, the \emph{`flavor-transition-couplings'} $
\mu_{\alpha[(y^{a_y})_{{}_{(m)}} j^{a_j}]}^{\beta[I^{\eta_{{}_I}}(y^{a_y})_{{}_{(m)}}]
 }$ in the sixth line of (\ref{unifyg71}) are determined by a total number of `$\vert\{\beta[I^{\eta_{{}_I}}(y^{a_y})_{{}_{(m)}}\}\vert -1$' uncorrelated \emph{maximally-free Gaussian-random parameters} $\{\kappa_{\alpha}(j^{a_j};(y^{a_y})_{(m)};I^{\eta_{I}})\}$, 
ensuring that they satisfy the following constraint, 
\begin{equation}\label{clr}
 \sum_{\beta[I^{\eta_{{}_I}}(y^{a_y})_{{}_{(m)}}]}^{\rm{all\;choices}}\; \mu_{\alpha[(y^{a_y})_{{}_{(m)}} j^{a_j}]}^{\beta[I^{\eta_{{}_I}}(y^{a_y})_{{}_{(m)}}]
 }\;\;\; =\;\; 1 
\end{equation}
One can consistently restore (\ref{aloha3}) from the generalized cascade operators. The reduction is consistent because the flavorless formulation developed in sections two, three and four is equivalent with the flavored formulation in this section, by restricting to only one purely-vertonic flavor and only one purely-relatonic flavor. Because of this, upon applying the summation-constraint (\ref{clr}) to the one-flavor case, we obtain only one coefficient $\mu_{\alpha}^{\alpha} = 1$, such that (\ref{unifyg71}) does consistently reduce to (\ref{aloha3}). Now, \emph{for better clarity}, and as the simplest example, we expose here the first-order cascade operators for the sub-theory of HoloQuantum HyperNetworks defined with flavorless vertons and with `$\pm$' flavored relatons. Here come all the non-identity cascade operators of the model, 
\begin{equation}\label{example}\begin{split} 
&\;\; \mathcal{C}_{j}^{I^{\eta_{{}_I}}} =\; \prod_{m=1}^{M}\;
\prod_{y_{(m)} \equiv\;\{y_1 ... \;y_m\}\;}^{{\rm all\;choices}}\;  
\mathcal{C}_{(-)}^{j \to I^{\eta_{{}_I}}}[y_{(m)}]\;\; 
\mathcal{C}_{(+)}^{j \to I^{\eta_{{}_I}}}[y_{(m)}]
\\
&\;\;\; \mathcal{C}_{(\pm)}^{j \to I^{\eta_{{}_I}}}[y_{(m)}]\; =\; (\; 1 - 
\breve{n}_{y_{(m)} j}^{\pm}\;)\; + \\
& \;\;\:\;\; + \big[\; \mu_{\pm}^{-}(j;y_{(m)};I^{\eta_{{}_I}})\;\breve{r}^{\dagger -}_{I^{\eta_{{}_I}} y_{(m)}}\; +\; \mu_{\pm}^{+}(j;y_{(m)};I^{\eta_{{}_I}})\; \breve{r}^{\dagger +}_{I^{\eta_{{}_I}} y_{(m)}} \;\big]\; \breve{r}_{y_{(m)}j}^{\pm} \\
& \;\mu_{\pm}^{\pm}(j;y_{(m)};I^{\eta_{{}_I}}) = \frac{1}{2} \;\pm\; \kappa_{\pm} (j;y_{(m)};I^{\eta_{{}_I}})\;\\
\end{split}\end{equation}
Let us confirm that principle five is indeed realized in the perfected theory. Generators of quantum-hypergraphical isomorphisms 
are given by the cascade operators in (\ref{unifyg71}),
\begin{equation}\label{vcp1811118}\begin{split}
& \Gamma_{j^{a_j}}^{i^{b_i}} \;\equiv\;  1 + n_j^{a_j} (1 - n_i^{b_i}) [\; v_j^{a_j}\; \mathcal{C}_{j^{a_j}}^{i^{b_i}}(\hat{\upmu}_{j^{a_j}}^{i^{b_i}} \;)\; v^{\dagger b_i}_i - 1\;] \\
& \hat{\upmu}_{j^{a_j}}^{i^{b_i}} \equiv \bigcup_{m = 1}^{M} \bigcup_{(y^{a_y})_{{}_{(m)}}}^{{\rm all\;choices}} \bigcup_{\alpha[(y^{a_y})_{{}_{(m)}} j^{a_j}]}^{{\rm all\;choices}} \bigcup_{\beta[i^{b_i}(y^{a_y})_{{}_{(m)}}]}^{{\rm all\;choices}}\; \{\; \delta_{\alpha[(y^{a_y})_{{}_{(m)}} j^{a_j}]}^{\beta[i^{b_i}(y^{a_y})_{{}_{(m)}}]}\; \}_{\vert_{(\ref{clr})}}  \end{split} \end{equation} By (\ref{vcp1811118}), \emph{the complete set of the quantum-hypergraphical isomorphisms in the perfected theory} are realized by the $\mathcal{C}_{j^{a_j}}^{i^{b_i}}$s whose flavor-transiotion-couplings are localized (by the tuned Gaussian distributions of their $\{\kappa\}$ variables) on `$\mu_\alpha^\beta = \delta_\alpha^\beta$', and in addition, they do satisfy (\ref{clr}). 
\\ Let us highlight the wisdom for having defined the cascade operators of the perfected theory as in (\ref{unifyg71}).  By the definitions in (\ref{unifyg71}), not only all the enlarged conversion operators $\breve{f}_J^{\gamma_{{}_{J}}} \mathcal{C}_{J^{\gamma_{{}_{J}}}}^{I^{\eta_{{}_{I}}}}\breve{f}_I^{\dagger \eta_{{}_{I}}}$ satisfy the flavored generalizations of (\ref{8abc}), and at the same time principle five is realized for the maxiamlly-flavored HoloQuantum HyperNetworks, but also the unitary dynamics of the perfected theory, as we now formulate, receives its most complete realization of all the nine principles.\\
%
\\ By straghtforwardly generalizing the results of sections three and four, and utilizing the maximally-flavored cascade operators given in (\ref{unifyg71}), we realize the \emph{ninth principle} to obtain as follows \emph{the total Hamiltonian of the  
Perfected HoloQuantum HyperNetwork Theory,} 
\begin{equation}\label{71h333333333h73}\begin{split}
& H^{(\mathcal{FM})}_{M} 
= 
\sum_{m=1}^{m=M^\star} 
H_M^{(\mathcal{FM})(m-m)} \\
& H_M^{(\mathcal{FM})(m-m)} = \\
& = \sum_{J_1 \cdot \cdot \cdot J_m}^{{\rm all}} \sum_{\gamma_{{}_{J_1}} \cdot \cdot \cdot \gamma_{{}_{J_m}}}^{{\rm all}}  \sum_{I_1 \cdot \cdot \cdot I_m}^{{\rm all}}  \sum_{\eta_{{}_{I_1}} \cdot \cdot \cdot \;\eta_{{}_{I_m}}}^{{\rm all}}
\lambda_{J_1^{\gamma_{{}_{J_1}}}\;\cdot \cdot \cdot\; J_m ^{\gamma_{{}_{J_m}}}}^{I_1^{\eta_{{}_{I_1}}}\; \cdot \cdot \cdot\;  I_m^{\eta_{{}_{I_m}}} }\; \Phi_{J_1^{\gamma_{{}_{J_1}}}\; \cdot \cdot \cdot\; J_m ^{\gamma_{{}_{J_m}}}}^{I_1^{\eta_{{}_{I_1}}}\;\cdot \cdot \cdot \;I_m^{\eta_{{}_{I_m}}}}\\ & \\ &  \hspace{.001 cm} \hspace{.00133 cm} \hspace{.00199 cm} \Phi_{J_1^{\gamma_{{}_{J_1}}}\; \cdot \cdot \cdot\; J_m ^{\gamma_{{}_{J_m}}}}^{I_1^{\eta_{{}_{I_1}}} \cdot \cdot \cdot  I_m^{\eta_{{}_{I_m}}}} 
= \; (\prod_{s = 1}^{m} \breve{f}_{J_{s}}^{\;\gamma_{{}_{J_{s}}}})\; \big(\mathcal{C}_{J_1^{\gamma_{{}_{J_1}}}}^{I_1^{\eta_{{}_{I_1}}}} \cdot \cdot \cdot\;\mathcal{C}_{J_m^{\gamma_{{}_{J_m}}}}^{I_m^{\eta_{{}_{I_m}}}}\big)\; 
(\prod_{r=1}^{m} \breve{f}_{I_r}^{\dagger 
\eta_{{}_{I_r}}}) \\ & \\
& \lambda_{J_1^{\gamma_{{}_{J_1}}}\;\cdot \cdot \cdot\; J_m ^{\gamma_{{}_{J_m}}}}^{I_1^{\eta_{{}_{I_1}}}\;\cdot \cdot \cdot\; I_m^{\eta_{{}_{I_m}}}} =  \lambda_{[J_1^{\gamma_{{}_{J_1}}}\;\cdot \cdot \cdot\; J_m ^{\gamma_{{}_{J_m}}}]}^{I_1^{\eta_{{}_{I_1}}}\;\cdot \cdot \cdot\; I_m^{\eta_{{}_{I_m}}}} = \lambda_{J_1^{\gamma_{{}_{J_1}}}\;\cdot \cdot \cdot\; J_m ^{\gamma_{{}_{J_m}}}}^{[I_1^{\eta_{{}_{I_1}}}\;\cdot \cdot \cdot\; I_m^{\eta_{{}_{I_m}}}]} \\
& \lambda_{J_1^{\gamma_{{}_{J_1}}}\;\cdot \cdot \cdot\; J_m ^{\gamma_{{}_{J_m}}}}^{I_1^{\eta_{{}_{I_1}}}\;\cdot \cdot \cdot\; I_m^{\eta_{{}_{I_m}}}} =  \bar{\lambda}_{I_1^{\eta_{{}_{I_1}}}\;\cdot \cdot \cdot\; I_m ^{\eta_{{}_{I_m}}}}^{J_1^{\gamma_{{}_{J_1}}}\;\cdot \cdot \cdot\; J_m^{\gamma_{{}_{J_m}}}} \\ &
\mathcal{P}(\lambda_{\vec{J}^{\;\vec{\gamma}_{\vec{J}}}}^{\vec{I}^{\;\vec{\eta}_{\vec{I}}}}\;|\;\mathring{\lambda}_{\vec{J}^{\;\vec{\gamma}_{\vec{J}}}}^{\vec{I}^{\;\vec{\eta}_{\vec{I}}}}\;,\;  \hat{\lambda}_{\vec{J}^{\;\vec{\gamma}_{\vec{J}}}}^{\vec{I}^{\;\vec{\eta}_{\vec{I}}}}) \;\sim\; \exp \big(\;- \frac{(\lambda_{\vec{J}^{\;\vec{\gamma}_{\vec{J}}}}^{\vec{I}^{\;\vec{\eta}_{\vec{I}}}} - \mathring{\lambda}_{\vec{J}^{\;\vec{\gamma}_{\vec{J}}}}^{\vec{I}^{\;\vec{\eta}_{\vec{I}}}})^2}{\hat{\lambda}_{\vec{J}^{\;\vec{\gamma}_{\vec{J}}}}^{\vec{I}^{\;\vec{\eta}_{\vec{I}}}}}\;\big) \end{split}\end{equation} 
By discerning all the degeneracies in the summations of  (\ref{71h333333333h73}), $H_M^{(\mathcal{FM})(m-m)}$ takes a detailed form similar to (\ref{ube111}), but now with all the flavors being incorporated. Now, knowing the complete unitary dynamics of the perfected theory, the \emph{`Compactified Maximally-Flavored HoloQuantum HyperNetwork Theory'} must be defined, similar to its flavourless counterpart in section four. This theory is the `minimal-maximal child sub-theory' of the perfected theory, with the same quantum statics as (\ref{equa71}), but with the total hamiltonian which exponentiates the first-degree Hamiltonian in (\ref{71h333333333h73}). Namely,
\begin{equation}\label{child99}\begin{split}
& \mathcal{H}_{M}^{(\mathcal{FM}_{\bigcirc})} \;=\; \mathcal{H}_{M}^{(\mathcal{FM})}\\ & 
\{\;\mathcal{O}^{(H)(\mathcal{FM}_{\bigcirc})}_{(m)}\;;\;\forall\;m_{\geq 1}^{\leq M^\star}\; \}\;=\;\{\;\Phi_{J_1^{\gamma_{{}_{J_1}}}\; \cdot \cdot \cdot\; J_m ^{\gamma_{{}_{J_m}}}}^{I_1^{\eta_{{}_{I_1}}} \cdot \cdot \cdot  I_m^{\eta_{{}_{I_m}}}} \;;\;\forall\;m_{\geq 1}^{\leq M^\star}\; \} \\
& H^{(\mathcal{FM}_{\bigcirc})}_{M} = \exp\big(\sum_{J\;,\;\gamma_{{}_{J}}}^{{\rm all}}\;\sum_{I\;,\;\eta_{{}_{I}}}^{{\rm all}}\; \lambda_{J^{\gamma_{{}_{J}}}}^{I^{\eta_{{}_{I}}}}\;
 \breve{f}_J^{\;\gamma_{{}_{J}}}\;
 \mathcal{C}_{J^{\gamma_{{}_J}}}^{I^{\eta_{{}_I}}}\;\; 
 \breve{f}_I^{\dagger \eta_{{}_I}}\big) 
 \end{split} 
 \end{equation} The perfected theory, which we have already concluded, has been axiomatically defined and formulated based on the nine principles. Most remarkably, its total Hilbert space (\ref{equa71}) and its 
complete unitary dynamics (\ref{71h333333333h73}) are, by their axiomatic constructions, as general, fundamental and complete as they can be. By the such-accomplished ultimate realization of principle nine, the perfected theory is covariantly complete. Being so, even when sheerly regarded as a theory of quantum mathematics, the Maximally-Flavored HoloQuantum HyperNetwork Theory is the \emph{kinematically-and-dynamically complete theory of `all possible forms' of mathematically-quantized hypergraphs}. Now, given the extremely-generic domain of all dynamical hypergraphs which are formulated by the flavorless theory, defined in the first four sections, we must take in what follows one and only one more step to work-out the `hypergraphical covariant completeness' of the maximally-flavored theory. That is, we must work-out how the perfected theory formulates all the dynamical mathematically-quantized hypergraphs which are defined with \emph{arbitrarily-oriented hyperlinks}, and with \emph{arbitrarily-weighted vertices and hyperlinks}.\\ 
\\ The natural and direct `it-from-qubit' way by which one presents the general, complete formulation of the maximally-oriented and maximally-weighted quantized hypergraphs is to have it extracted as a sub-theory of (\ref{equa71}) and (\ref{71h333333333h73}), upon defining all the orientations and all the weights as specific \emph{flavors}. Let us state the whole point. \emph{In HoloQuantum HyperNetwork Theory, every relaton of every arbitrarily-chosen orientation or weight, and every verton of every arbitrarily-chosen weight, must be distinctively taken to be one independent quantum degree of freedom of} (\ref{equa71}), \emph{that is, it should be identified with one formally-fermion qubit} $F_I^{\gamma_{{}_{I}}}$ \emph{which is addressed with a `correspondence-flavor'}. Being so, in the promised theory of the maximally-oriented and maximally-weighted HoloQuantum HyperNetworks, one can `weight stuff' and `orient stuff' just with the formally-fermionic qubits. Restated, the complete set of relatonic orientations and weights, and their complete set of vertonic weights must be mapped in one-to-one manners to their coresspondence sets of 
flavors.\\
\\ Let us now begin to formulate the above-featured sub-theory of the perfected $\mathcal{M}$ with identifying all the hyperlink orientations in HoloQuantum HyperNetworks. By definition, a hyperlink of $m$ base-vertices has $m!$ orientations, each one of them corresponding to a unique ordering of its base-vertices. As such, to formulate the most general orientations of the quantum hyperlinks, every $m$-relaton, identified earlier in (\ref{equa7}) as $\breve{R}_{i_{(m)}}$, must be now `branched-out' as the $m!$ `orientationally-flavored relatons' $\breve{R}_{i_{(m)}}^{\alpha_{(m)}}$ in the Hilbert space (\ref{equa71}). Namly, \emph{the maximally-oriented HoloQuantum hypergraphs} must be formulated by this family of \emph{orientation flavors},
\begin{equation}\label{orientations} \forall\;i_{(m)}\;\;\;,\;\;\;
\breve{R}_{i_{(m)}}\;\;\; \hookrightarrow \;\;\; \{\; \breve{R}_{i_{(m)}}^{\alpha_{(m)}}\;\;;\;\alpha_{(m)} \in \mathbb{N}^{\leq m!}\;\}
\end{equation}
Like orientations (\ref{orientations}), all the \emph{hypergraphical weights} form a distinct set of the flavors, named the `\emph{weight flavors}'. Stating precisely, the `all-inclusive set' of all the defined weights for the structural constituents of a HoloQuantum hypergraph, being vertices or the $m$-degree hyperlinks, is mapped in a one-to-one manner to a complete set of vertonic-and-relatonic weight flavors. Being maximally general, the weight flavor of every structural-qubit $F_I$ comes with its distinct labeling $\omega(F_I) \equiv \omega_I$, and also with a distinct arbitrarily-chosen spectrum $\mathcal{W}_I \equiv \{\omega_I\}$. That is every hypergraph-state-qubit $F_I$ is branched-out by taking flavors as follows,
\begin{equation}\label{weights} \forall\;I\;\;\;;\;\;\; F_I \;\;\; \hookrightarrow \;\;\; \{\;F_I^{\;\omega_I}\;\;\;;\;\;\;\omega_I \in \mathcal{W}_I\}\;\;\;;\;\;\; \end{equation}
Now, the total quantum kinematics and the complete unitary quantum dynamics of all possible HoloQuantum HyperNetworks which are \emph{both maximally oriented and maximally weighted} are given by the Hilbert space (\ref{equa71}) and the total Hamiltonian (\ref{71h333333333h73}) whose  \emph{`structural flavors'} are identified as follows,
\begin{equation}\label{orientationsandweights} \{\; F_I^{\gamma_{{}_I}} \;\} \;\equiv\;\{\; V_i^{\omega_i}\;\;\;;\;\;\; R_{(i^\omega)_{(m)}}^{\alpha_{(m)};\omega[{(i^\omega)_{(m)}]}}\;\}  \end{equation} We highlight a point about the many-body interactions (\ref{71h333333333h73}) of HoloQuantum HyperNetwork Theory endowed with the structural flavors (\ref{orientationsandweights}). This point is a direct implication of the maximally-flavored version of the fundamental rule, which is equal to the very same rule stated in section three, upon the replacements $F_I \hookrightarrow F_I^{\gamma_{{}_{I}}}$ and $\mathcal{C}_J^I \hookrightarrow \mathcal{C}_{J^{\gamma_{{}_{I}}}}^{I^{\gamma_{{}_{I}}}}$. Given this rule, at the level of the fundamental formulation of the theory, not only the quantum vertices and all the quantum hyperlinks of different degrees can freely convert into one another, but also now all their different orientations and all their different weights can convert into one another.\\
\\ The triplet (\ref{equa71},\ref{71h333333333h73},\ref{orientationsandweights}) does conclude the promised mathematical covariant-completion of the flavorless `basic-class' theory, by defining and formulating the HoloQuantum HyperNetworks which are endowed with general orientations and weights. That is, \emph{the kinematically-and-dynamically complete `it-from-qubit' formulation of all possible mathematically-quantized hypergraphs is the above `sub-theory triplet'} (\ref{equa71},\ref{71h333333333h73},\ref{orientationsandweights}).\\ \\
But, now as an instructionally-good example of the HoloQuantum HyperNetworks whose some of flavors take continuous spectra, and also because weights are usually taken to be real-or-complex valued numbers, \emph{let us treat the spectra of the weight-flavors to be the arbitrarily-chosen continues number-fields}. By this take, the defining operators of the hypergraph-state-qubits will be turned into formally-fermion field-function operators living on the topological product of the time-dimension and an abstract `wight-space'. Being so, one will obtain a \emph{`HoloQuantum-HyperNetworkical Continuum Quantum Field Theory'} living on this abstract space, albeit at a merely formal level.\\ To this aim, we now represent even-handedly all the hypergraph-state-qubits which are, as already specified, both maximally oriented and maximally weighted by the field-function operators $\breve{f}_I^{(\dagger)\alpha_I}(\vec{\omega}_I)\;\equiv\;\breve{f}_I^{(\dagger)\alpha_I}(\vec{x}_I)$, where,
\begin{equation}\label{hqncqft}
\{\breve{f}_I^{(\dagger)\alpha_I}(\vec{x}_I)\} 
= \{\;v_i^{(\dagger)}(\vec{x}_i)
\;;\;
r_{i_{(m)}}^{(\dagger)\alpha_{(m)}}
(\vec{x}_{i_{(m)}}\;|\;\vec{x}_1 , \cdot \cdot \cdot, \vec{x}_m)\;\}
\end{equation} 
In the right hand side of (\ref{hqncqft}), each finite-dimensional vector $\vec{x}_I$ identifies the continuously-valued weight of the corresponding hypergraph-state-qubit $F_I$, whose spectrum can be independently chosen, to be maximally general. For example, if the corresponding weight is a real number, $\vec{x} \equiv x^{\;\in \mathbb{R}}$, but if it is a complex number, $\vec{x} \equiv (z,\bar{z})^{\;
\in \Sigma}$ with $\Sigma$ being the complex plain $\mathbb{C}$ or any Riemannian surface whose topology one chooses arbitrarily. So, in the models where all the weights are real numbers, or complex numbers, we obtain a formally-all-fermionic HoloQuantum-HyperNetworkical Quantum Field Theory, with novel interactions, living on abstract Lorentzian manifolds similar to the $(1+1)$ dimensional or to the $(2 + 1)$ dimensional spacetimes. To highlight, HoloQuantum HyperNetworks with multiple hyperlinks correspond to the special case in which relatons are integrally weighted, namely
$\vec{x} = n^{\in \mathbb{N}}$.\\\\ 
The cascade operators $\mathcal{C}_{J^{\alpha_{{}_J}}}^{I^{\beta_{{}_I}}}(\vec{x}_J;\vec{x}_I)$  and the total Hmailtonian of the complete unitary dynamics of this continuously-flavored sub-theory are defined as in (\ref{unifyg71},\ref{71h333333333h73}), upon turning the  summations-and-products over all the continuous weights into integrations and the exponentials of logarithmic integrations, respectively. In particular, its \emph{compactified sub-theory} is defined by,
\begin{equation}\begin{split}\label{child9}&H^{(\mathcal{FM}_\bigcirc)}_{M} = \exp\big(\;\sum_{J,I} \sum_{\alpha_J,\beta_I} \int d\vec{x}_I d\vec{x}_J\; \lambda_{J^{\alpha_J}}^{I^{\beta_I}}(\vec{x}_J;\vec{x}_I)\;\times\\ &\hspace{1.99 cm}\hspace{.85 cm}\hspace{.33 cm}\;\;\;\times\;\breve{f}_J^{\alpha_J}(\vec{x}_J)\;\mathcal{C}_{J^{\alpha_J}}^{I^{\beta_I}}(\vec{x}_J;\vec{x}_I)\; \breve{f}_I^{\dagger \beta_I}(\vec{x}_I)\;\big)
\end{split}\end{equation}
Results (\ref{equa71},\ref{71h333333333h73}) conclude the `Perfected HoloQuantum HyperNetwork Theory', $\mathcal{M}$, axiomatically built upon all the nine principles. As promised in section one, this is as such the fundamental and the complete form-invariant `it-from-qubit' theory of both all the dynamical quantum hypergraphs and all quantum natures. 
\section{HoloQuantum HyperNetwork Theory: The Models With Extra Symmetries, Supersymmetric Models} 
As HoloQuantum HyperNetwork Theory is built to be the quantum many body theory of all physically-possible systems of quantum objects and quantum relations, it \emph{must be necessarily maximally minimalistic in taking its fundamental symmetries}. This `minimalism' is implied by the very nine principles, so that the only symmetries transformations in $\mathcal{M}$ are \emph{`the unavoidable ones'}.\\
Being maximally minimalistic in taking fundamental symmetries, $\mathcal{M}$ only has the exact $U(1)$ symmetry of the global-phase redundancies, the minimally-broken symmetry of qubits-equal-treatment, and has developed within it the complete set of quantum-hypergraphical isomorohisms by specific combinations of a subset of its many-body-interaction operators. On the other hand, in the contextually-developed or the domain-specific sub-theories and models, solutions or phases of $\mathcal{M}$, generically additional \emph{exact-or-approximate symmetries}, either must be placed merely as the \emph{extra-structure `constraint impositions'}, or must be emerged as the \emph{extra-structure `emergencies'}. We highlight here two extremely-motivated classes of such extra symmetries. Because hypergraphs are, as fundamentally defined, pregeometric and background-less, there can not be any spatially-local fundamental symmetry or
fundamental gauge redundancy in the perfected theory. However, in both the \emph{`emergent-spacetime'} and \emph{`standard-model-like'} sub-theories or phases of $\mathcal{M}$, a number of appropriate gauge redundamcies must take place as such extra symmetries, in either of the above ways.\\\\
As an instructive and also interesting illustration of this point, here we will incorporate \emph{supersymmetry} as one such \emph{extra symmetry} in a maximal sub-theory of the perfected theory. As it indeed must be, it will be demonstrated that to formulate the most general family of supersymmetric HoloQuantum HyperNetworks, one needs neither to introduce any extra degrees of freedom into the total quantum kinematics, nor to deform the complete total quantum dynamics of the perfected theory. That is, $\mathcal{M}$ stays kinematically and dynamically form-invariant under this extra-symmetry imposition, although indeed some of its otherwise-free structures will become constrained. We formulate here two different model-theories of HoloQuantum HyperNetworks which are 
endowed with $\mathcal{N}=2$ global supersymmetries. Moreover, to illustrate the simplest yet sufficiently-rich examples, we suffice here to the symmersymmetrization of the HoloQuantum HyperNetworks which are flavorless and in which only relatons are kept quantumly-active. The recipes will yield us supersymmetric HoloQuantum HyperNetworks which are formally-purely-fermionic and whose Hamiltonians are embedded in (\ref{h111be11}).\\\\ One must pick up a(ny possible) complex-conjugate pair of the formally-fermionic operators $(Q_M^{(\mathcal{M}_{(s)})},  \bar{Q}_M^{(\mathcal{M}_{(s)})})$ on the Hilbert space $\mathcal{H}_{M}^{(\mathcal{M})}$, which being supercharges, turn every formally-fermion hypergraph-state qubit $\breve{f}^{(\dagger)}_I$ into a \emph{formally-composite-boson qubit} $\breve{b}^{(\dagger)}_I$ as its supersymmetric partner, and vice versa. The algebra is, 
\begin{equation}\label{susybridge}
\begin{split}
& (Q_M^{(\mathcal{M}_{(s)})})^2 = (\bar{Q}_M^{(\mathcal{M}_{(s)})})^2 = 0\\ 
&[Q_M^{(\mathcal{M}_{(s)})}, \breve{f}^\dagger_I] = \breve{b}_I^{\dagger }\;;\;[Q_M^{(\mathcal{M}_{(s)})},\breve{f}_I] = 0 \;;\;h.c\\  
&[Q_M^{(\mathcal{M}_{(s)})}, \breve{b}_I] = \partial_t \breve{f}_I\;;\;[Q_M^{(\mathcal{M}_{(s)})},\breve{b}_I^{\dagger }] = 0 \;;\; h.c
\end{split}
\end{equation}
We formulate \emph{a very general family of supersymmetruc HoloQuantum HyperNetworks}. By being so, the defining supersymmetric charges $(Q_M^{(\mathcal{M}_{(s)})},\bar{Q}_M^{(\mathcal{M}_{(s)})})$ must be the \emph{`largest-possible' Gaussian-random composite fermionic fields}, being made from the qubits $\{\breve{f}_J\};\{\breve{f}_I^{\dagger}\}$,  
in such a way that the supersymmetric Hamiltonian, $H_{M}^{(\mathcal{M}_{(s)})}$, 
\begin{equation}\label{susyh1}
H_{M}^{(\mathcal{M}_{(s)})} \equiv \{Q_M^{(\mathcal{M}_{(s)})},\bar{Q}_M^{(\mathcal{M}_{(s)})}\}
\end{equation} 
satisfies the \emph{`maximal-dynamical-embedding'} criterion,
\begin{equation}\label{susyh2} H_{M}^{(\mathcal{M}_{(s)})}\; \subset^{{}^{({\rm embedded\;maximally})}}\; \mathcal{H}_{M}^{(\mathcal{M})}
\end{equation}
The first supersymmetrization recipe is inspired by the initiative work \cite{BM16NWJ}, being the best method of realizing \emph{a supersymmetric sub-theory of} $\mathcal{M}$ \emph{which is maximal both kinematically and dynamically}. By this first recipe, one imposes an entirely-arbitrary $\mathbb{Z}_2$ partitioning on the total Hilbet space $\mathcal{H}_{M}^{(\mathcal{M})}$. Imposing so, the  formally-fermion hypergraph-state qubits $\breve{F}_I$ are partitioned arbitrarily into two `chirality' classes,
\begin{equation}\label{ssc}\begin{split}
&\{\;\breve{f}^{(\dagger)}_I\;\}\;\equiv\;\{\;\psi_{\hat{I}(-)}^{(\dagger)}\;;\;\psi_{\hat{I}(+)}^{(\dagger)}\;\}\\
&\;\;\; \mathbb{Z}_2\;:\;\; \psi_{\hat{I}(-)}^{(\dagger)}\; \longleftrightarrow\; \psi_{\hat{I}(+)}^{(\dagger)} \end{split}\end{equation}
Let us highlight that the choice of this $\mathbb{Z}_2$ partitioning is absolutely arbitrary, and so all the different choices for it results in the supersymmetric models which may differ interpretationally, but indeed as supersymmetric quantum theories are all physically equivalent. Next, in accordance to the chirality partitioning (\ref{ssc}) imposed on the hypergraph-state-qubits $\breve{F}_I$, the two supersymmetric charges of the sub-theory $\mathcal{M}_{(s)}$ are so defined,
\begin{equation}\label{bsc}\begin{split}
&Q_{M}^{(\mathcal{M}_{(s)})} \equiv\;
\sum_{\hat{J}_1}^{{\rm all}}\; \eta_{\hat{J_1}}  \psi_{\hat{J_1}(+)} + \sum_{\hat{J}_1 \hat{J}_2 \hat{I}_1}^{{\rm all}}\; \eta_{\hat{J_1} \hat{J_2} \hat{I}_1} \psi_{\hat{J_1}(+)} \psi_{\hat{J_2}(+)} \psi_{\hat{I_1}(-)}^{\dagger} + . . . =\\
&\hspace{.99 cm} \hspace{.33 cm} \hspace{.66 cm} = \sum_{m=0}^{\frac{M^\star}{2}-1}\;\sum_{\hat{J}_{(m+1)}}^{{\rm all}}\;\sum_{\hat{I}_{(m)}}^{{\rm all}}\;
\eta_{\hat{J}_1 \cdot\cdot\cdot \hat{J}_{m+1} \hat{I}_1 \cdot \cdot \cdot \hat{I}_m}\;\times\\
&\hspace{1.99 cm}\times\;(\psi_{\hat{J_1}(+)} \cdot \cdot \cdot \psi_{\hat{J}_{(m+1)}(+)})\;(\psi_{\hat{I}_m(-)}^\dagger \cdot \cdot \cdot \psi_{\hat{I}_1(-)}^\dagger)\\
&\bar{Q}_{M}^{(\mathcal{M}_{(s)})} \;\equiv\;
\sum_{\hat{I}_1}^{{\rm all}} \;\bar{\eta}_{\hat{I}_1}\; \psi_{\hat{I}_1(+)}^\dagger + \sum_{\hat{J}_1 \hat{I}_1 \hat{I}_2}^{{\rm all}} \bar{\eta}_{\hat{J}_1 \hat{I}_1 \hat{I}_2}\; \psi_{\hat{J}_1(-)} \psi_{\hat{I}_2 (+)}^\dagger \psi_{\hat{I}_1 (+)}^\dagger  + ... =\\
&\hspace{.99 cm}\hspace{.66 cm} \hspace{.33 cm} = \sum_{m=0}^{\frac{M^\star}{2}-1}\;\sum_{\hat{J}_{(m)}}^{{\rm all}}\;\sum_{\hat{I}_{(m+1)}}^{{\rm all}}\;
\bar{\eta}_{\hat{J}_1 \cdot \cdot \cdot \hat{J}_{m} \hat{I}_1 \cdot\cdot\cdot \hat{I}_{m+1}}\;\times\\
&\hspace{1.99 cm}\times\;(\psi_{\hat{J}_1(-)} \cdot \cdot \cdot \psi_{\hat{J}_m(-)})\;(\psi_{\hat{I}_{(m+1)}(+)}^\dagger \cdot \cdot \cdot \psi_{\hat{I}_{1}(+)}^\dagger) 
 \end{split}\end{equation}
In the definitions (\ref{bsc}), all the independent defining couplings $\eta_{\hat{J}_1 \cdot\cdot\cdot \hat{J}_{m+1};\hat{I}_1 \cdot \cdot \cdot \hat{I}_m}$s are the statistically uncorrelated Gaussian random parameters being selected the same distributions (\ref{h445}) for the $\lambda$-couplings of $\mathcal{M}$.\\
By selecting the supersymmetric charges to be the ones in (\ref{bsc}), as we can indeed compute,
one now realizes the maximal sub-theory of (flavorless and purely-relatonic) supersymmetric HoloQuantum HyperNetworks, $\mathcal{M}_{(s)}$. That is, the resulted supersymmetric total Hamltonian (\ref{susyh1}) does satisfy the dynamical maximality criterion (\ref{susyh2}), by developing a family of \emph{`composite-random couplings'},
\begin{equation}\label{susyh3}\begin{split}
& (\lambda_{J_1 \cdot \cdot \cdot J_m}^{I_1 \cdot \cdot \cdot I_m}\;;\; \bar{\lambda}_{J_1 \cdot \cdot \cdot J_m}^{I_1 \cdot \cdot \cdot I_m}\;) =\;\\ & =\; {\rm some\;c.c\; functions\;of}\; \{\;\eta_{\hat{J}_1 \cdot \cdot \cdot \hat{J}_{s} \hat{I}_1 \cdot\cdot\cdot \hat{I}_{s+1}}\;,\;\bar{\eta}  _{\hat{J}_1 \cdot \cdot \cdot \hat{J}_{r} \hat{I}_1 \cdot\cdot\cdot \hat{I}_{r+1}}\; \} \end{split}
\end{equation}
We highlight that such a `$(\pm)$ partitioning' of the formally-fermionic Hilbert space is very natural in many of the sub-theories and models which are derived from within $\mathcal{M}$, an example of which being \emph{a Wheelerian simplest toy model} presented in the next section. Being so, all such models are naturally amenable to the supersymmetrization presented above, albeit whenever their model-defining constraints respect supersymmetry. \\\\ Let us here mention an alternative recipe. Similar to the supersymmetrization in \cite{BM16SYKsusy}, the supercharges can be also defined as follows, 
\begin{equation}\begin{split}\label{scs} &Q_M^{(\mathcal{M}_{(s)})}\; \equiv\; \sum_{m=1}^{\frac{M^\star}{2}} 
q_{(2m-1)} \sum_{{\rm all}\;Js}^{1 \leq s \leq m} \eta_{J_1 \cdot \cdot \cdot J_{2m-1}}  \breve{f}_{J_1}\cdot \cdot \cdot 
\breve{f}_{J_{2m-1}} \equiv \\
&\;\;\;\hspace{.111cm}\;\;\;\;\;\;\;\;\;\equiv \sum_{m=1}^{\frac{M^\star}{2}}\;q_{(2m-1)}\;Q_{M (2m-1)}^{(\mathcal{M}_{(s)})}\;\;;\;\;\;
\bar{Q}_{M}^{(\mathcal{M}_{(s)})} \equiv \big(Q_{M}^{(\mathcal{M}_{(s)})}\big)^\dagger
\end{split}\end{equation}
with the $q_{(2m-1)}$s being formal complex coefficients, by which all the `fixed-order supercharges' $Q_{M (2m-1)}$ are distinguished. By taking the supercharges given in (\ref{scs}), the resulted supersymmetric Hamiltonian (\ref{susyh1}) develops, in addition to the resulted embedding of $H_{M}^{(\mathcal{M})}[\{\lambda_{\vec{J}}^{\vec{I}}\}, c.c,\}]$, the following terms $\Delta H_M$, 
\begin{equation} \begin{split} & \Delta H_M = \sum_{0 \leq m,n \leq \frac{M^\star}{2}}^{(n \neq m)} \sum_{\vec{J}_{(2m)},\vec{I}_{(2n)}}^{{\rm all\;possible}} \lambda_{\vec{J}_{(2m)}}^{\vec{I}_{(2n)}} (\prod_{L}^{{\rm some}} \breve{n}_{L}) (\prod_{s=0}^{2m} \breve{f}_{J_s}) (\prod_{r=0}^{2n} \breve{f}_{I_r}^\dagger)\\ &\hspace{1.55cm}\lambda_{\vec{J}_{(2m)}}^{\vec{I}_{(2n \neq 2m)}}\; \equiv q_{(2m+1)} \bar{q}_{(2n+1)} \sum_{K}^{M^\star}  \eta_{K \vec{J}_{(2m)}} \bar{\eta}_{K \vec{I}_{(2n \neq 2m)}}\end{split}\end{equation} However, these extra terms do violate the global $U(1)$ symmetry of HoloQuantum Network Theory, which is necessary for its well-definedness as a background-less quantum theory.  By demanding the preservation of the $U(1)$ symmetry, $\Delta H_M = 0$, we get a set of algebraic constraints on the $\eta$ parameters, whose generic solution allows only a single $q$-coefficient to be non-zero. By this, we will get a sub-theory with only a pair of `fixed-order supercharges' $(Q_{M (2m-1)}^{(\mathcal{M}_{(s)})}\:; h.c)$. As such, we favor the first supersymmetrization recipe in light of the principle nine of $\mathcal{M}$.
\section{HoloQuantum HyperNetwork Theory:
\hspace*{.00077 cm} A Simplest Toy Model of `Wheelerian Participatory Universe'}
In this section, we work-out from within the Perfected HoloQuantum HyperNetwork Theory a specific `simplest toy model' on two purposes. \emph{First}, we want to expose how concrete models of phenomenological interests can be extracted-out from the total quantum kinematics and the complete unitary quantum dynamics of the theory $\mathcal{M}$. \emph{Second}, we wan to present a minimalistic simplest toy model of the \emph{`HoloQuantum-HyperNetworkically Realized Wheelerian Quantum Universe'}. By being Wheelerian, we mean a quantum universe which is constructed upon, and so realizes, \emph{all of the `three' visions of Wheeler} \cite{BM16JW}. These three principal visions are so stated. \emph{First}, the total quantum universe is fundamenally a gigantic quantum [Hyper-]Network of the 
\emph{`many-observers-participancies'}, whose `actions' are her `elementary quantum phenomena'. \emph{Second}, there is a fundamental maximal statistical randomness underlying this quantum HyperNetwork, sourcing the \emph{`law(s) without law(s)'}. \emph{Third}, the principle of \emph{`it from [qu-]bit'} states that \emph{absolutely all the physical aspects of the world} is assembled by the randomly-sourced `answering-qubits' of `no-or-yes's to the abundantly-many binary questions posed by all the participatory-observers. We emphasize that the manners in which these three principles are incorporated in the toy model here are intentionally set to be the most minimalistic, the simplest, and sheerly instructive. But, here comes the one single significant massage that we want to convey here. \emph{The above three Wheelerian principles can be merged-and-realized correctly by `the theory $\mathcal{M}$' which has been defined, formulated and finally perfected in this work. Being so, the `no-or-yes' qubits of the observer-participatory universe must be identified with the very qubits (\ref{equa71}) which construct HoloQuantum HyperNetworks and so microscopically interact with one another as determined by the total Hamiltonian of the perfected $\mathcal{M}$ theory} 
(\ref{71h333333333h73}). \\\\ Let us now conceive and formulate step-by-step the simplest toy model of a Wheelerian participancy universe which is `purely-relational'. By being purely relational, we mean the following simplification in this minimal toy model. We will take all of the participatory observes to be fixedly frozen in their states of presence, and then let the relatonic qubits of the `multi-observer participany relations' be the only dynamical quantum degrees of freedom. It is clear that in the ultimate theory of the Wheelerian universe, every participatory observer must also be quantumly active as a dynamical verton which can switch between the states of absence and presence. This implies that all the nontrivial cascade operators should be necessarily in role in any realistic Wheelerian model. However, in here we reduce the today model to be purely relational, on the account of the first purpose of this section which is instructional.\\\\
By this explanation, let us consider a total number of $M$ 
participatory observers who are all fixedly present. To formulate their Wheelerian Participatory Universe from within $\mathcal{M}$, let us employ the total Hilbert space $\mathcal{H}_{M}^{(\mathcal{FM})}$ (\ref{equa71}) of $M$ flavorless vertons together with all their `evenly-flavored' $m$-relatons. The vertons of $\mathcal{H}_{M}^{(\mathcal{FM})}$ represent, by any one-to-one correspondence, those very $M$ quantum participatory observers.  Hence, by our simplifying ansatz, all the vertons $V_i$ are frozen in their presence-states, in a way that must be consistent with the dynamics of the system. Because they are fixedly present, the first set of the quantum constraints of the toy model are the following operator identities, 
	\begin{equation}\label{oc1}
		n_i = 1\;\;\;,\;\;\,\forall\; i\: \in \mathbb{N}^{\leq M}
	\end{equation}
Wheelerianly, \emph{the fundamental degrees of freedom} which microscopically \emph{assemble} this whole quantum universe, are \emph{all the randomly-sourced `no-or-yes' answering-qubits to a complete set of questions posed by, and shared by, all the subsets of all the $M$ participatory-observers}. The \emph{completeness} of the set of binary questions means that if we were given the qubit data about any proper subset of those questions, the entire participatory universe could not have been assembled only upon them. Now, we let every complete set of these `no-or-yes' answering-qubits be collected in the set of doublets $\{(\mathcal{N}_{i_{(m)}}^{\alpha_{i_{(m)}}}\; ,\; \mathcal{Y}_{i_{(m)}}^{\alpha_{i_{(m)}}})\}$, in which $i_{(m)}$ identifies the corresponding choice of $m^{\leq M}$ participatory questioners, while the index $a_{i_{(m)}}$ forms \emph{a complete spectrum} of the binary questions posed by the observers. The simplest HoloQuantum HyperNetworkian toy-modeling of this Wheelerian scenario is `all relatonic'. Therefore,  the hypergraph-state relatons should not only carry the very same question-spectrum flavors, but also be further doubly-flavored, as follows,
\begin{equation}\label{rays}\begin{split}  &\{\;\breve{R}_{i_{(m)}}^{\alpha_{i_{(m)}} , \nu}\;\} \;\equiv\; \{\;\breve{R}_{i_{(m)}}^{\alpha_{i_{(m)}} , \pm}\;\}\\ &\{\; \breve{R}_{i_{(m)}}^{\alpha_{i_{(m)}}, - }\;\} \longleftrightarrow \{\;\mathcal{N}_{i_{(m)}}^{\alpha_{i_{(m)}}}\;\}\\ & \{\; \breve{R}_{i_{(m)}}^{\alpha_{i_{(m)}}, + }\;\} \longleftrightarrow \{\;\mathcal{Y}_{i_{(m)}}^{\alpha_{i_{(m)}}}\;\} \end{split}\end{equation} 
By (\ref{rays}), every such-identified degree-$m$ relaton whose chirality flavor is positive (negative) represents the positive (negative) answer given to one of the binary questions posed by those $m$ participatory observers who are in correspondence with its base-vertons. Let us highlight that, by (\ref{oc1}), the hypergraph-state-relatons $\breve{R}_{i_{(m)}}^{\alpha_{i_{(m)}} , \nu}$ are operationally equal to their counterprarts  $R_{i_{(m)}}^{\alpha_{i_{(m)}} , \nu}$, everywhere in this simplest toy model.\\
\\\ Relatons (\ref{rays}) are the active quantum degrees of freedom which assemble the Wheelerian Universe. Namely, the operators $\big(\breve{r}_{i_{(m)}}^{\alpha_{i_{(m)}} , \pm} , \breve{r}_{i_{(m)}}^{\dagger \alpha_{i_{(m)}} , \pm}\big)$ are set free to annihilate and create their `no-or-yes' answering-qubits, and so, every new instant \emph{to re-assemble the universe}.\\\\
Because of the abstract purely-information-theoretic `binary' nature of every pair of relatons with opposite chiralities, one must impose one more set of quantum constraints. \emph{The defining basis-states of the total Hilbert space must be mapped, in a one-to-one manner, to the
outcomes of the `all-measurements' which correspond to the `complete-questions-answered' entirely-assembled Wheelerian universes of the} $M$ \emph{participatory observers}. By this demand, the evolving global wavefunction of the Wheelerian universe must be, at any arbitrary time, a superposition of the basis-states in which, for every choice of ${m^{\leq M}}$ observers $i_{(m)}$, and for every of the items $\alpha_{i_{(m)}}$ in their questionnaire, one of the two $R_{i_{(m)}}^{\alpha_{i_{(m)}},\pm}$ relatons is `present', whereas its chirality-complement one is necessarily `absent'. Formulated as operator identities, we must now demand the following second set of quantum constraints as relatonic operator identities,
	\begin{equation}\label{oc2}
	\breve{n}_{i_{(m)}}^{\alpha_{i_{(m)}},-} +
	\breve{n}_{i_{(m)}}^{\alpha_{i_{(m)}},+}  = 1\;\;\;,\;\;\;\forall\; i_{{(m_{\leq M})}}\; ,\; \forall\: \alpha_{i_{(m_{\leq M})}} \end{equation}
By imposing (\ref{oc1}) and (\ref{oc2}), the total Hilbert space of the simplest toy model of the \emph{Wheelerian Participatory Universe of the} $M$ \emph{observers}, $\mathcal{H}_{M}^{(\mathcal{W.P.U})}$, is given by the truncation of $\mathcal{H}_{M}^{(\mathcal{FM})}$ whose defining basis is identified as follows,
\begin{equation}\label{ast}\begin{split}
 & \mathcal{B}_M^{(\mathcal{W.P.U})} \;=\;  \hspace{.00111cm} \hspace{.01 cm} \hspace{.007 cm} \{\;\prod_{m = 1}^{M}\;\prod_{i_{(m)}}^{{\rm all}}\;\prod_{\;\alpha_{i_{(m)}}}^{{\rm all}}\; \big[\;
 (1 - \epsilon_{i_{(m)}}^{\alpha_{i_{(m)}}})\; \breve{r}^{\dagger \alpha_{i_{(m)}} , -}_{
 	i_{(m)}}\;
+ \\ & \hspace{.33 cm} \hspace{.33 cm} \hspace{.33 cm} \hspace{1.99 cm} \hspace{.66 cm}\;\;\;\;\;\;\;\;\;\;\;\;\hspace{.5 cm}+\; \epsilon_{i_{(m)}}^{\alpha_{i_{(m)}}}\; \breve{r}^{\dagger \alpha_{i_{(m)}} , +}_{
i_{(m)}}\;\;
 \big]
 \ket{0}
\\&\hspace{.033 cm}\hspace{.99 cm} \hspace{1.099 cm}\;\;\;\;\;\;\;,\;\;{\rm for\;all\;possible\;choices\;of}\; \epsilon_{i_{(m_{\leq \hat{M}})}}^{\alpha_{i_{(m)}}} \in \{0,1\}
\;\}\hspace{1 cm} 
\end{split} \end{equation} Moreover, the quantum-kinamatical truncation (\ref{ast}), which in turn is demanded by the quantum constraints (\ref{oc1},\ref{oc2}), must be dynamically consistent. That is, clearly, the above kinematical truncation must be preserved in the entire evolution of these Wheelerian HoloQuantum HyperNetworks. For this \emph{dynamical consistency} to come true, the complete total Hamiltonian of the Wheelerian Participatory Universe, $H_{M}^{(\mathcal{W.P.U})}$, which generates its unitary evolution, must satisfy the following costraints,
\begin{equation}\label{art123}
\begin{split}
&[\;H_{M}^{(\mathcal{W.P.U})}\;,\; n_i \;] \; =\; 0\;\;\;,\;\;\;\forall\;i_{\leq M}\\
&[\;H_{M}^{(\mathcal{W.P.U})}\;,\;\breve{n}_{i_{(m)}}^{\alpha_{i_{(m)}},-} +
	\breve{n}_{i_{(m)}}^{\alpha_{i_{(m)}},+}  \;] \; = \; 0\;\;\;,\;\;\;\forall\;i_{(m_{\leq M})}\;,\;\forall\; \alpha_{i_{(m)}} \end{split} \end{equation} 
Being promised from the very beginning, the above total Hamiltonian $H_{M}^{(\mathcal{W.P.U})}$ must be \emph{directly} extracted from the total Hamiltonian of the perfected $\mathcal{M}$, namely (\ref{71h333333333h73}). To fulfill this, we take the first-order Hamiltonian of the perfected theory $\mathcal{M}$, manifested in the exponent of (\ref{child99}), and have it now expressed as the likewise-flavored cousin of the flavorless unfoldment (\ref{insideh11}).
\\ By applying (\ref{art123},\ref{oc1},\ref{oc2}) to  that, and dropping constant terms, we present as follows the \emph{first-degree Hamiltonian of the `most minimalistic' toy model of the HoloQuantum HyperNetworkical Wheelerian Participatory Universe},
\begin{equation}
\begin{split}\label{bwumh11} 
& H_{M}^{(\mathcal{W.P.U})(1-1)}\;=\; \hspace{.11 cm}  \sum_{{\rm all}\;i_{(m)}}^{1 \leq m \leq M} \sum_{{\rm all}\;\alpha_{i_{(m)}}} \big(\;\;\mu_{i_{(m)}}^{\alpha_{i_{(m)}}, +}\; \breve{n}_{i_{(m)}}^{\alpha_{i_{(m)}},+}\;\;+\hspace{.11 cm}\\ & \hspace{3.33 cm} +\;\;\;\bar{\lambda}_{i_{(m)}}^{\alpha_{i_{(m)}},-+}\;\; \breve{r}_{i_{(m)}}^{\alpha_{i_{(m)}},+}\; \breve{r}_{i_{(m)}}^{\dagger \alpha_{i_{(m)}},-}\; +\\ & \hspace{3.33 cm} +\;\;\; \lambda_{\;i_{(m)}}^{\alpha_{i_{(m)}},-+}\;\;\breve{r}_{i_{(m)}}^{\alpha_{i_{(m)}},-}\; \breve{r}_{i_{(m)}}^{\dagger \alpha_{i_{(m)}},+}\;\;\big)\hspace{.03 cm}\; \end{split} \end{equation} Let us highlight that, because in this simplest model of Wheelerian HoloQuantum HyperNetwork, all the vertons are kept present fixedly by the quantum constraint (\ref{oc1}), the nontrivial cascade operators do not appears in the Hamiltonian. To remind, all the independent $\mu_{i_{(m)}}^{\alpha_{i_{(m)}},+}$ and the $\lambda_{\;i_{(m)}}^{\alpha_{i_{(m)}},-+}$ couplings in (\ref{bwumh11}) must be taken to be uncorrelated maximally-free Gaussian-random as in (\ref{71h333333333h73}). The above first-degree Hamiltonian can be more compactly re-expressed 
as such,
\begin{equation}\label{gh1} H_{M}^{(\mathcal{W.P.U})(1-1)} = \sum_{i_{(m_{|_{m_{\geq 1}^{\leq M}}})}}^{{\rm all}}\;\sum_{\alpha_{i_{(m)}}}^{{\rm all}}\; \sum_{\nu,\upsilon}^{\in \{\pm\}}\;
\lambda_{i_{(m)}}^{\alpha_{i_{(m)}},\nu \upsilon}\;\mathcal{T}_{i_{(m)};\nu \upsilon}^{\;\alpha_{i_{(m)}}(1-1)}\;
\end{equation}
using the so-defined \emph{`Elementary Wheelerian Operators'},
\begin{equation}\label{ewo}
\hspace{.0199cm}\;\;\;\;\;\mathcal{T}_{i_{(m)};\nu \upsilon}^{\;\alpha_{i_{(m)}}(1-1)} \;\equiv\;\;\; \breve{r}_{i_{(m)}}^{\alpha_{i_{(m)}}, \nu}\; \breve{r}_{i_{(m)}}^{\dagger \alpha_{i_{(m)}}, \upsilon}
\end{equation}
As the indices manifest, the above operators $\mathcal{T}_{i_{(m)};\nu \upsilon}^{\;\alpha_{i_{(m)}}(1-1)}$ are defined for every subset of the participatory-observers $i_{(m)}$, and further, for every one $\alpha_{i_{(m)}}$ in the complete spectrum of their binary questions. By definition, for every such identification, the elementary Wheelerian operators either switch the two answering-chiralities, the off-diagonal ones, or simply witness-and-report the chiralities, the diagonal ones, in the present-moment global state of the purely `it-from-qubit' universe.\\\\
Now, upon utilizing the expressions (\ref{gh1},\ref{ewo}), the complete total Hamiltonian of the simplest toy model of the Wheelerian Participatory Universe $H_{M}^{(\mathcal{W.P.U})}$, takes the following form, as a sub-evolution of (\ref{71h333333333h73}), 
\begin{equation}\label{wuwei}
\begin{split}
&H_{M}^{(\mathcal{W.P.U})}\;=\; \sum_{s}^{1 \leq s \leq M}\;H_{M}^{(\mathcal{W.P.U})(s-s)}\\
&H_{M}^{(\mathcal{W.P.U})(s-s)} \;\equiv\;\\&\equiv\;
(\sum_{{\rm all}\;{i_1}_{(m_1)}}^{1 \leq m_1 \leq M}\;\sum_{{\rm all}\;\alpha_{{i_1}_{(m_1)}}}\;\sum_{\nu_1,\upsilon_1}^{\in \{\pm\}}) \;\cdot \cdot \cdot\; (\sum_{{\rm all}\;{i_s}_{(m_s)}}^{1 \leq m_s \leq M}\;\sum_{{\rm all}\;\alpha_{{i_s}_{(m_s)}}}\; \sum_{\nu_s,\upsilon_s}^{\in \{\pm\}})\\
&\hspace{.113366 cm}\lambda_{{i_1}_{(m_1)} \cdot \cdot \cdot {i_s}_{(m_s)}}^{\alpha_{{i_1}_{(m_1)}} \cdot \cdot \cdot \alpha_{{i_s}_{(m_s)}},\nu_1\upsilon_1\; \cdot \cdot \cdot \;\nu_s\upsilon_s}\;\mathcal{T}_{i_{1_{(m_1)}};\nu_1\upsilon_1}^{\;\alpha_{{i_1}_{(m_1)}}(1-1)}  \cdot \cdot \cdot \;\mathcal{T}_{i_{s_{(m_s)}};\nu_s\upsilon_s}^{\;\alpha_{{i_s}_{(m_s)}}(1-1)} 
\end{split}
\end{equation}	
Being one of the central characteristics of the unitary evolution (\ref{wuwei}), the `no-or-yes' answering-qubits  $\breve{R}_{i_{(m_{\leq M})}}^{\alpha_{i_{(m)}} , \pm}$ do all interact with one another as conducted by the maximally-random many-body conversion operators of the theory $\mathcal{M}$.\\ 
\\ To complete this special section, let us also present the \emph{`child' sub-model of this minimal toy model of the HoloQuantum-HyperNetworkically-Realized
Wheelerian Participatory Universe}. As before, it is the one whose total Hamiltonian exponentiates (\ref{bwumh11}), \begin{equation}\begin{split}\label{S}
&H_{M}^{(\mathcal{W.P.U})_\bigcirc} \;=\; \exp \big[\; \sum_{{\rm all}\;i_{(m)}}^{1 \leq m \leq M} \sum_{{\rm all}\;\alpha_{i_{(m)}}} \sum_{\nu,\upsilon}^{\in \{\pm\}}\;
\lambda_{i_{(m)}}^{\alpha_{i_{(m)}}, \nu \upsilon}\; \mathcal{T}_{i_{(m)};\nu \upsilon}^{\;\alpha_{i_{(m)}}(1-1)} \;\big]
\end{split}\end{equation}
\section{HoloQuantum HyperNetwork Theory: A Global Overview With\\ Selected Connectional Explanations} We begin this section with  \emph{a global overview} of the theory $\mathcal{M}$, and then bring about a number of \emph{selective} connectional remarks which are important.\\\\ HoloQuantum HyperNetwork Theory, which we have defined and systematically formulated in this work, is the fundamental kinematically-and-dynamically complete theory of all the entirely-quantized HyperNetworks. From the purely \emph{physics} point of view, $\mathcal{M}$ serves the very job which \emph{the complete `it-from-qubit' theory of all physics} should do. To the best of our understanding, as we propose here, HoloQuantum HyperNetwork Theory is indeed the fundamental complete `it-from-qubit' quantum-many-body form-invariant formulation of `all quantum natures' in an all-unified way. We remember that, `all quantum nature' is a collective term for the entire quantum universe-or-multiverse and for all her `selective descendants' one by one, as obtained by all possible `phenomenological subsettings' together with all possible `observers-probes rescalings'. All the quantum kinematics, all the many-body interactions, and the complete unitary quantum dynamics of the perfected theory $\mathcal{M}$ are directly sourced from the \emph{unique nine principles} which, to the best of our undedrstanding, are both the unavoidable, `the must be', and the most compelling,   
`the best be', for its original intention to be entirely fulfilled.\\\\\ 
Mathematically, the central characters of $\mathcal{M}$ are the HoloQuantum Hypergraphs, the unitarily-evolving quantum states which are made by all possible quantum superpositions of the arbitrarily-chosen hypergraphs. But, the microscopic degrees of freedom of $\mathcal{M}$ are the complete sets of purely-information-theoretic abstract qubits for the quantum vertices and for the quantum hyperlinks, the vertons and the relatons, respectively.\\ As a whole, all the vertons-and-relatons form an entirely pregeometric, `formally'-all-fermionic, \emph{qualitatively-novel} closed quantum many body system of abstract qubits. The total unitary qauntum dynamics of all the vertons and relatons is sourced by a complete set of random many-body interactions which impartially occur in between all of them. Besides all the `conventional'  multi-fermion interactions, \emph{HoloQuantum HyperNetwork Theory has a whole lot of novel many-body interactions, being conducted by a hierarchical family of the `cascade operators'}, as implied by the evolving hypergraphical face of its abstract quantum many body system of qubits. These novel and cardinal cascade operators safeguard the dynamical hypergraphical-weldefinedness of HoloQuantum HyperNetworks, but also realize (by only the purely-vertonic subset of them) all the quantum hypergraphical isomorphisms.\\\\ 
Both mathematically and physically, all that the theory $\mathcal{M}$ `takes in' about the nature of the vertons and the relatons is their Wheelerian abstract `no-or-yes' or equivalently `absence-or-presence' qubit-ness for a complete set of the maximally-flavored vertices and their multi-degree hyperlinks. Being so, $\mathcal{M}$ \emph{is purely `it-from-qubit' in its fully-covariant formulation of both mathematics and physics, all the way from `alpha to omega'}. Because of this central characteristic, and moreover because of its \emph{covariant-completeness}, the perfected theory yields the complete formulation of how precisely all the abstract `no-or-yes' qubits of the observer-participancies interact with one another and evolve as an abtract quantum many-bpdy system, assembling the Wheelerian universes and multiverses. As such, as an alternative to approaches such as \cite{BM16Wen}, $\mathcal{M}$ \emph{is proposed as the right theory to develop in full precision the `it-from-qubit' construction of all physics}.\\\\ To accomplish this, HoloQuantum hypergraphs represent the arbitrary choices of the quantum objects, by their vertons, and the arbitrary choices of the multi-object quantum relations, by their $m$-relatons. Indeed, the purely relatonic operators $\mathcal{O}[\{\breve{r}_{(i^{a_i})_{{}_{(m)}}}^{\alpha[(i^{a_i})_{{}_{(m)}}]},\breve{r}_{(i^{a_i})_{{}_{(m)}}}^{\dagger\;\alpha[(i^{a_i})_{{}_{(m)}}]}\}]$s, being correctly identified in every given context in terms of the fundamental relatons, can be the field operators of the arbitrarily-chosen physical interactions between the arbitrarily-chosen `particles'. Likewise, all the statistical correlations, the functional relations, or the relational geometrical quantifications of these `particles' can be formulated by these operators. In fact, all possible fundamental-or-emergent relational observables, which one can measure for any arbitrarily-chosen `particles', can be formulated by these operators. However, it is only by inter-relating those very `objects', that the relational observables are even definable. By taking-in all these objects as the qubit vertons, one arrives at $\mathcal{M}$.  \emph{One so covariantly formulate all of physics by HoloQuantum HyperNetworks of arbitrary objects-and-relations}.\\ 
Having made a review of HoloQuatum HyperNetwork Theory, we now come to present, in the rest of this section, \emph{a number of elucidating connectional remarks about `some selected works' in the literature which has some `instructional' overlaps with some aspects of the perfected theory} $\mathcal{M}$. In each case, we will comparatively comment and carefully elaborate not only on the notable conceptual or technical connections and similarities, but also on the multi-dimensional characteristic distinctions which HoloQuantum HyperNetwork Theory makes with all those selected works. The works which we comparatively discuss in here belong to very different fields of quantum physics. This vast coverage, however, is very natural as the perfected theory is by construction the form-invariant formulation of all quantum natures. \emph{The following selected works from the literature}, \emph{we must highlight, are picked up very `subjectively', mainly for the `instructional' purposes, and so their collection is `far from complete'}. This highlighted selectiveness is mainly because the selected works are already enough for clariyfing the points. These remarks also shed light on how $\mathcal{M}$ can advance all these fields.\\\\
Mathematically, the absolute primitivity, maximal generality, and intrinsic competence which distinguishes the defining framework of hypergraphs, suggest them to be the very primary characters of the \emph{`pregeometry'} \cite{BM16PG} out of which the entire space emerges. Graphs, namely hypergraphs with only the two-vertex-links, have a well-known history of being examined as models of pregeometry. Because the entire universe, 
and so also the fundamental `setting' of the spacetime, is a quantum system, the hypergraphical pregeometry must be defined quantumly. Indeed, some interesting quantum models of graphical pregeometries have been already constructed in the past and recent literature \cite{BM16A, BM16DP, BM16MR, B1MLP, BMR161MHLKCSLM,BM16GB,BM16CAT}. These works, independent of the implemented quantum statistics of the degrees of freedom, \emph{differently} but all \emph{partially}, have employed some features of the total quantum many body system of the theory $\mathcal{M}$. Both of the works \cite{BM16A, BM16DP} already used the framework of second quantization for the graph-structural degrees of freedom, in \cite{BM16DP} 
with some deterministic local evolution, and in \cite{BM16A} with some random interactions. The more recent and more advanced works of \cite{BMR161MHLKCSLM}, looking for emergent locality and gravity, formulate initiative models of second-quantized quantum-graphical pregeometries, also including matter quanta, whose in-particular Hubbard-model-resembling structures-and-interactions develop interesting phases. The works of \cite{BM16GB}, using a second-quantized formalism being similar to the quantum-graphical methods of \cite{BMR161MHLKCSLM}, present models of random complexes whose Markovian state-dependent unitary evolutions are realzied by gluing face-wise the simplices or the regular polytopes. Finally, the most recent works \cite{BM16CAT} offer simple toy models of the randomly-interacting graph-structural qubits whose ground-states, developed in the infrared quantum phase transitions, feature some four-diemsnional geometries.  
\\ Let us now elaborate collectively in what follows on the significant points of distinctions between what we have presented in this work and those presented in \cite{BM16A, BM16DP, BM16MR, B1MLP, BMR161MHLKCSLM,BM16GB,BM16CAT}, as both the partial similarities between them and their partial overlaps must be already obvious to the reader.\\\\ \emph{First}, HoloQuantum HyperNetwork Theory is \emph{a complete `theory'}.  By intention, this theory is constructed to serve as the fundamental and complete theory by which `all of physics', and so also the complete theory of quantum pregeomtry and quantum gravity, can be formulated form-invariantly. Because it is neither a specific model, nor a collection of contextually-related models, its definition and formulation could not have been done in the arbitrary manners in which `models' are built. Being so, nothing in this quantum theory is ad hoc, or has been implemented as a matter of examination, simplification, or specifications. Both the total statics and the complete unitary dynamics of the theory, as they must, are directly deduced from the merge of the stated nine principles required by the  `must-be'-and-`best-be'. This characteristic robust-and-complete determination is unlike all the ad-hoc structures implemented in the initiative models presnted in \cite{BM16A, BM16DP, BM16MR, B1MLP, BMR161MHLKCSLM,BM16GB,BM16CAT}. To highlight, all the many-body interactions of the theory are determined both uniquely and completely by the exact-or-almost symmetries, all the dynamical quantum constraints, the Wheelerian randomness, and finally the principle of the covariant-completeness. Indeed, \emph{these very microscopic interactions}, upon realizing all the extra structures or the extra symmetries which are needful to be imposed in obtaining the correct theory of quantum gravity from within HoloQuantum HyperNetwork Theory, \emph{will also formulate form-invariantly the complete dynamics of the quantum spacetiem}. We will elaborate on this crucial point in section nine.\\\\ \emph{Second}, we must remark here on a characteristic distinction regarding the inclusion of all the additional fundamental degrees of freedom  which will play the role of matter in quantum pregeometry and quantum gravity. On one hand, $\mathcal{M}$ is a complete `purely internal theory', without any external degrees of freedom. HoloQuantum HyperNetworks are all the quantum states of a closed quantum many body system of the abstract qubits of vertons and relatons which formulate the total kinematics and the complete dynamics of the fully-quantized hypergraphs. On the other hand, by partitioning this total quantum many body system into two complementary subsystems, one formulates `open HoloQuantum Networks'. Such complementary pairs of open HoloQuantum Networks naturally formulate `the spacetime quanta' interacting with `the matter quanta' in the aimed theory of quantum gravity. Being so, \emph{the vertons and $m$-relatons of HoloQuantum HyperNetworks can formulate both the `pure quantum geometries' and the `interacting quantum spacetimes and all possible quantum matter', in a covariant `unified' way}.\\ \emph{Third}, one independent dimension in which any HoloQuantum-HyperNetworkical theory of the quantum pregeometry and quantum gravity must be significantly distinct statically, by interactions and dynamically from the models in \cite{BM16A, BM16DP, BM16MR, B1MLP, BMR161MHLKCSLM,BM16GB,BM16CAT}, is the following. The theory $\mathcal{M}$ not only is `trans-graphical', but also both kinematically and dynamically is `maximally hypergraphic'. This is indeed as it must be, and is directly sourced by the realization of principle seven, according to which \emph{all the} $m^{> 2}$-\emph{relatons, which by definition are `non-reducible' to the two-relatons, are as kinematically-fundamental and also as dynamically-active as the two-relatons}. Stated by principle seven, the theory must treat all the hypergraph-state-qubits with maximum-possoible equality. The minimalistic breaking of this symmetry is between the vertons and the relatons, only realized by the `dressing' of relatons and by the cascade operators. Besides this, all qubits are equal. As such, by its total Hamiltonian, namely $H_M^{(\mathcal{FM})}$, all the $R_{i_{(m\;\in \mathbb{N}^{\leq M})}}$s are equally activated. As we understand, this  quality of `\emph{all-relatons-equality}', being an exact fundamental symmetry of the theory, is instrumental for the correct theory of quantum gravity. Let us so conclude as follows. \emph{The acceptable HoloQuantum-Networkical formulations of quantum pregeometry and quantum gravity are the ones which are `maximally hypergraphic'}. Comparing $H_M^{(\mathcal{FM})}$ $(\ref{71h333333333h73})$ with its counterparts in the aforementioned references proves the physical significance of this point. \\\\ \emph{Finally}, let us conclude our connectional remarks on the models of quantum spacetime with implications of \emph{the covariant completeness} stated by principle nine. Because both the total state-pace (\ref{equa71}) and the complete unitary dynamics (\ref{71h333333333h73}) of the perfected theory do fully realize this principle, it is already guaranteed that `all' the network models whose kinematics and dynamics are `quantumly correct' can be consistently embedded inside the totality of the perfected $\mathcal{M}$. These consistent embeddings can be also checked concretely, example by example. In particular, \emph{all the models presented in} \cite{BM16A, BM16DP, BM16MR, B1MLP, BMR161MHLKCSLM,BM16GB,BM16CAT} \emph{are consistently embeddable inside} (\ref{equa71},\ref{71h333333333h73}). To highlight, as explained in sections one to five and in the overview part of this section, this `consistent-embedding criterion' holds independent of the quantum-statistical identities of the objects or relations whose `it-from-qubit' theory is formulated by the perfected theory. \emph{That, all the `independent' models in} \cite{BM16A, BM16DP, BM16MR, B1MLP, BMR161MHLKCSLM,BM16GB,BM16CAT} \emph{are embeddable inside the Maximally-Flavored HoloQuantum HyperNetwork Theory not only interconnects them fundamentally, but also makes all their consistent deformations and generalizations manifest}.\\\\
Now, we move on to the comparative remarks in regard with the SYK models \cite{BM16SYK, BM16SYKearlier,BM16SYKGR}, 
specially 
with 
their `complexified 
versions' 
\cite{BM16SYK}. On one hand, an inclusive comparison will be made briefly. On the other hand, \emph{this highlights how one can generate the qualitatively
novel classes of quantum many body systems, inside} $\mathcal{M}$.\\
\emph{Merely at the level of their formulations as quantum many body systems}, HoloQuantum HyperNetworks have both a number of notable \emph{likenesses} and a number of significant characteristic \emph{unlikenesses} with the extremely interesting SYK models, specially with their complexified formulations as given in \cite{BM16SYK}. We begin by highlighting \emph{the `formal' likenesses}. The total Hilbert spaces are purely fermionic, in both cases. All the microscopic interactions, in both cases, are fundamentally  `all-to-all' and also random. Moreover, simply for the naturalness, but not restrictedly, the random couplings are all set to be Gaussian. Both HoloQuantum HyperNetwork Theory and the `complexified SYK' as formulated in \cite{BM16SYK} respect the fundamental symmetry of the global $U(1)$ transformations on the fundamental fermions.  Finally, in both cases, the formulations are totally pregeometric. That is, the fermionic many body systems live in $(0+1)$ spacetime dimensions. We must emphasize again that these likenesses are at a \emph{merely formal level}.\\\\ Let us now move on to highlight and elaborate briefly on \emph{the characteristic unlikenesses} which, \emph{even merely formulationally}, distinguish the works. We so begin with `\emph{the least significant}' among these differences. Although in both cases the fundamental all-to-all interactions are all Gaussian-random, in HoloQuantum HyperNetwork Theory, the independent couplings \emph{must be} selected from the Gaussian-random distributions each one of which is fixed with tuning two arbitrary parameters, one for the mean values, and one for the standard deviations. Both (formally) in HoloQuantum HyperNetwork Theory and in the model \cite{BM16SYK} together with the likewise-complexified extended-SYK model \cite{BM16SYKGR}, the interactions are the multi-fermionic-conversions. But, the total Hamiltonian of HoloQuantum HyperNetwork Theory $H_M^{(\mathcal{FM})}$ in (\ref{71h333333333h73}) \emph{`must necessarily' contain all} the $(m$-to-$m)$ conversions impartially, to realize the \emph{covariant completeness} stated by principle nine.\\\\ Now, we turn to remarking \emph{the three unlikenesses} between the works which are `\emph{the most significant}'. \emph{First}, in the SYK models \cite{BM16SYKearlier,BM16SYK,BM16SYKGR} all the fermionic degrees of freedom are treated equally, and this equal treatment which is both statical and dynamical, is held exactly. But, in HoloQuantum HyperNetwork Theory this global `equal-treatment symmetry' must be explicitly broken, as minimally as possible, but indeed unavoidably. The source of the explicit breaking of this symmetry is \emph{a fundamental one}. In HoloQuantum HyperNetwork Theory, the qubits $F_I$  fundamentally sit into two categories, vertons and relatons. Relatons can be present or created only when their base-vertons are all present in the quantum state defining the HyperNetwork. Being so, relatons are `system-state-conditional qubits'. By principle seven, the equal-treatment symmetry is broken only by these relatonic conditionalities. \emph{But, as} (\ref{71h333333333h73}) \emph{does manifest in comparison to} \cite{BM16SYK, BM16SYKearlier, BM16SYKGR}, \emph{this effect strongly deforms the unitary dynamics} $H_M^{(\mathcal{FM})}$.\\ \emph{Second}, because by seeing on its very mathematical face, the maximally-flavored HoloQuantum HyperNetwork theory `becomes equivalent with' the most complete theory of the all-structurally-quantized dynamical hypergraphs, the \emph{quantum-hypergraphical isomorphisms} do play an important role in the construction of its dynamical side, as stated by principle five. Indeed, realizing these quantum-isomorphism transformations has made a significant impact on the definitions and the formulations of the microscopic interactions and the total unitary evolution of the perfected theory, as one can trace them back in the first five sections. But, in the SYK models \cite{BM16SYK,BM16SYKearlier,BM16SYKGR} the relatonic structures are all trivial, so that this dynamically-impactful feature becomes effectively mute.\\\\ \emph{Third}, the dynamical hypergraphical-welldefinedness of HoloQuantum HyperNetworks demands the hierarchical family of \emph{cascade operators}. These novel operators play a central role in the fundamental interactions and so in the unitary evolution of the total many body system of the vertons and all the $m$-relatons. However, because the hypergraph structure of the SYK models is effectively trivial, \emph{the interactions of HoloQuantum HyperNetwork Theory} \emph{and of the models} \cite{BM16SYK,BM16SYKearlier,BM16SYKGR} \emph{are majorly distinct at the microscopic level}, because of the both highly-frequent and highly-impactful presences of the cascade operators in $H_M^{(\mathcal{FM})}$ as concluded in (\ref{71h333333333h73}).\\\\ 
Beyond the above-stated distinctions, the theory $\mathcal{M}$ is, by construction, the fundamental complete theory in which \emph{every} consistent quantum many body system is covariantly contained. \emph{In regard with the SYK-type models, in fact, all these embeddings are very straightforward}. That is, not only the models \cite{BM16SYK,BM16SYKearlier,BM16SYKGR} are immediately embeddable in the Maximally-Flavored HoloQuantum HyperNetwork Theory, but also all the descendant versions of them which are presented in the very recent literature, whether are generalized dimensionally, or are deformed statically or dynamically, 
can be easily embedded inside the very totality of (\ref{equa71},\ref{71h333333333h73}). By embedding all these SYK-type models into HoloQuantum HyperNetwork Theory, \emph{all their consistent deformations and generalizations becomes manifest, interconnected and systematic}.\\\\
Finally, we highlight a point on the gravitational aspects of the SYK models \cite{BM16SYKearlier,BM16SYK}, connecting with our related points in section nine. The SYK models are already dual to some two-dimensional gravitational models. But, the perfectly-realized principle nine does guarantee that there must be a \emph{sub-theory} of the theory $\mathcal{M}$ which holographically \cite{BM16RBMSH} defines and formulates the complete realistic theory of both quantum and classical gravity in the emergent realistic spacetimes in which, besides the single `renormalization-group dimension', the other dimesnions are all emergent. 
\newpage \section{HoloQuantum HyperNetwork Theory: The Visions Forward,\\ Today And Future}
HoloQuantum HyperNetwork Theory, understood in \emph{its `minimum level'}, serves the whole quantum-granted physics the very same way that \emph{category theory} serves the whole mathematics. But, it does serves physics \emph{much more than that}, once being understood in \emph{its `maximum level'}. The theory $\mathcal{M}$, by its definition, formulation and perfection, given the unique choice of its defining nine principles, serves as the right `it-from-qubit' framework in which every \emph{precedented-or-unprecedented} specific theory of physics can be constructed from the beginning, and then be built-up phenomenologically in a systematic way which is \emph{direct, fundamental and also optimal}. Being visited in the perspective of network science, HoloQuantum HyperNetwork Theory is \emph{`the' theory} which formulates every possible dynamical network, assuming that it respects all the laws of quantum physics structurally and functionally, which it surely does if being physically realizable. \\\\ 
In a way, it is  by those unique nine principles which HoloQuantum HyperNetwork Theory has been granted its intentional power, that is, the above \emph{`minimum and maximum'}. The nine principles are characterized into two classes which are `complementary'. \emph{Six principles} among them, the \emph{principles one, two, four, five, six and finally nine} are unavoidable for its minimal realization as \emph{`the category theory of all physics'}. Being so, they belong to the `must-be' class of these nine principles. In particular, we must highlight that, although the systematic construction of the perfected theory has been impacted, according to the principles five and six, by the realizations of the two distinct types of symmetry transformations, none of them is a `beyond-categorical' feature. Clearly, to develop any quantum-hypergraphic categorical theory of arbitrary objects-and-relations, the most complete set of the quantum-hypergraphical isomorphisms must be realized by a (proper) subset of the Hamiltonian operators. Moreover, the $U(1)$ redundancies of the global phases must be demanded for any `lowest-dimensional' categorical theory of quantum objects and their quantum relations to compute its observables correctly.
Now, we come to the second class of the nine principles, namely the `best-be'. They are the principles \emph{seven, three and eight}. Principle seven, namely the principle of `maximal hypergraphness' by which all the multi-degree quantum hyperlinks are equally-treated, and moreover all the quantum objects and the quantum relations are also treated as equally as it can be, is needed to make the perfected theory \emph{`the optimal framework'} in which all the domain-specific theories of physics can be formulated. Finally, the Wheelerian principles three and eight are the ones which \emph{uplift HoloQuantum HyperNetwork Theory to its ultimate `it-from-qubit' fulfillment} \cite{BM16JW}. \\ 
Summarized from this purely `it-from-qubit' point of view, HoloQuantum HyperNetwork Theory is the fundamental and complete dynamical interacting theory of the abstract qubits for the absences-or-presences of  `absolutely whatever' of the objects and their relations in the entire quantum universe or the multiverse. 
Clearly, the complete time-dependent information of `all quantum natures' is capturable by the total quantum many body system of a qualitatively sufficiently-diverse and quantitatively sufficiently-immense collection of the answring-qubits to the `isn't-or-is' or equivalently, to the `no-or-yes' questions. This is why, by its first-principle definition and perfected construction, the theory $\mathcal{M}$ \emph{is the fundamental, general, complete and covariant Wheelerian theory of `it-from-qubit'}.  This, in particular, suggests that from within the prerfected theory, one can generate a whole families of novel more compelling sub-theories and models of quantum information and quantum computation. We will suffice to highlight three obvious directions in this fruitful territory. \emph{Firstly}, one should be able to \emph{reformulate and further genralize both statically and dynamically, the conventional quantum computation theory}, from within the totality of (\ref{equa71}) and (\ref{71h333333333h73}). \emph{Secondly}, but relatedly, based on the results of \cite{BM16ACUQC} which are embeddable in, and generalizable by the total abstract microscopic system of HoloQuantum HyperNetwork Theory, one can devise \emph{highly-novel quantum computation processors which can be superior functionally}. \emph{Thirdly}, by hugely completing the simplest toy-model of section seven, one derives from within $\mathcal{M}$  \emph{the complete Wheelirian `It-From-Qubit' theory of `Observer-Participancy-Universe'}, an alternative to \cite{BM16Wen}.\\\\
HoloQuantum HyperNetwork Theory \emph{must be} maximally minimalistic in taking its fundamental symmetries to safeguard its maximal generality as the fundamental quantum many body theory of \emph{all} quantum natures. But, one can devise diverse models of HoloQuantum HyperNetworks endowed with the contextually-chosen extra symmetries. These contextual model buildings can be done in two ways. On one hand, as in the examples of section six, 
they can be obtained as the minimalizations being directly extracted from the total theory which has been fully perfected in section five. On the other hand, many forms of physically-significant extra symmetries, such as the spacetime symmetries or the internal gauge symmetries, can also emerge as the qualitatively-distinct quantum-or-classical phases of HoloQuantum HyperNetworks in the enormous total phase diagram of the perfected $\mathcal{M}$. Also from a purely network-science viewpoint, in these very two ways one can devise all models of unprecedented-or-precedented quantum-or-classical `simple'-or-complex networks. By its clear significance for building all such sub-theories and models, to probe and progressively map \emph{the distinct phases, the fixed points, and the phase transitions of the Maximally-Flavored HoloQuantum HyperNetworks} will be one of the fruitful directions to explore in future.\\
As elaborated in section eight, the theory $\mathcal{M}$, by being the fundamental complete `it-from-qubit' theory which formulates the time evolutions of all possible superpositions of the arbitrarily-structured hypergraphs, is the very natural framework to define, formulate and conclude the ultimate theory of \emph{quantum pregeometry and the complete quantum gravity}. To accomplish this far-reaching goal from within the totality of (\ref{equa71},\ref{71h333333333h73}) which conclude the perfected theory, is of course a whole grand project to be conducted, \emph{surely requiring a number of totally-unprecedented ideas}. To advance in the correct direction toward this goal, we suffice to highlight here the one most-significant central characteristic which is already very well-appreciated. This distinctive feature is nothing but \emph{`the correct form' of the most complete realization of Holography} \cite{BM16RBMSH}, one must come up with. This intrinsics holographization is clearly one more remark on models of section eight. Let us highlight that the whole point in here is \emph{the correct way} of realizing the strongest version of Holography, formulated in its purely-information-theoretic form and then implemented \emph{as an `extra' feature or structure on the whole generality of HoloQuantum HyperNetworks}. Namely, the \emph{`HoloQuantum-HyperNetworkian Theory of Quantum Gravity'} will be \emph{a very specific `sub-theory' of} 
the total HoloQuantum HyperNetwork Theory which, by intrinsic construction,  is immensely larger than a theory of only quantum gravity. This unmapped dimension must be explored in
future 
works.\\\\
Being one \emph{context-independent} direction of analysis within the HoloQuantum HyperNetwork Theory, one can systematically and also precisely analyze \emph{both the emergences and the characterizations of complexities} in a unique manner which is all-inclusive. By this we mean that, all the variants of complexities which are distinct phenomenologically, or are developed out of the different emergent mechanisms, can be unitedly formulated, classified and thoroughly understood in this unique framework. Given the extremely significant roles which \emph{the complex structures} do play both in the nature and in advanced technologies, this will be a distinctly-important field of study. In particular, because the theory $\mathcal{M}$ is the fundamental, general and complete unification of (quantum and classical) many body systems and (quantum and classical) networks, one can now attempt to systematically interconnect the defining characteristics, the foundational principles, and the emergent mechanisms of complexities \emph{in complex systems in physics and in complex networks}. Besides, by modelling \emph{all} the computationally-distinct classes of the computation processors as function-specific types of HoloQuantum HyperNetworks, one can reformulate unitedly and analyze more optimally \emph{computational complexities in both mathematics and computer science}. Finally, the theory $\mathcal{M}$, by being scale-covariant (on which we will elaborate next), is ideal for the study of the emergences of complexities even `practically'.
\\ The theory $\mathcal{M}$, by realizing all of its nine principles, is a quantum many body theory whose total kinematics, all microscopic interactions, and total unitary evolution do remain form-invariant in formulating all quantum natures. By the operationally constructive definition of `all quantum natures' given in section one, and also as explicitly stated in the explanatory note to the principle nine, this implies the following important statement. HoloQuantum HyperNetwork Theory is form-invariant under changing the renormalization group scale all the way down from the ultraviolet fixed point. Therefore, the quantum equations of the perfected theory are all `scale-covariant'. Being so, one knows that \emph{proposing HoloQuantum HyperNetwork Theory as the fundamental complete form-invariant `it-from-qubit' theory of all quantum natures is not to be meant only in a `scale-wise' conventional top-down manner}. Namely, although we can limit the theory to function `scale-by-scale', and so only in fixed scales, it can do much better in functioning over the renormalization-group scales. Because of its realized covariant-completeness, \emph{the perfected} $\mathcal{M}$ \emph{is intrinsically the `Multi-Scale Theory' in which quantum objects-and-relations in the arbitrarily-chosen different renormalization-group scales can become all activated simultaneously and cooperate with one another}. Given the proven significance of the multi-scale-functioning complex systems both in nature and in the advanced technologies, it remain as an important mission for the future works \emph{to manifestly formulate and thoroughly analyze, both at structural and functional levels, all the `multi-scale organizational aspects' of the theory} $\mathcal{M}$.\\\\ 
HoloQuantum HyperNetwork Theory, we must highlight, is being proposed in the present work as \emph{the complete covariant `it-from-qubit' theory of all quantum natures at the fundamental level, but not as `the ultimate theory'}. By this, one brings to attention that quantum physics is taken for granted in HoloQuantum HyperNetwork Theory, namely as a fundamental and exact input in it. But, it may be that some defining aspects of quantum physics, or even all of them, are emergent form a still-unknown `pre-quantum theory'. We suggest that the related problem of `effectively undoing time' in HoloQuantum HyperNetwork Theory, which we highlight in the conclusive paragraph of this paper, may serve as a good theoretical laboratory to experiment with the initiative models of the pre-quantum theory.\\\\ 
Finally, we must come to the very notion and the very role of \emph{`time'} in the perfected theory $\mathcal{M}$. As a quantum theory, time plays the role of the `external' evolution parameter in HoloQuantum HyperNetwork Theory. But, we may be able to \emph{`internalize time' relatonically} into the abstract total microscopic body of HoloQuantum HyperNetworks. By this, we will be led to even \emph{a one-level-higher parental theory}. 

\section*{Acknowledgments}
Alireza Tavanfar would like to very happily thank Yasser Omar, and moreover, Wissam Chemissany, Bruno Coutinho, Istvan A. Kovacs, Filippo Miatto, Masoud Mohseni, Giuseppe Di Molfeta, Mohammad Nouri Zonoz, Ali Parvizi, Marco Pezzutto, Daniel Reitzner, Alberto Verga and Mario Ziman for very precious discussions. Alireza Tavanfar also thanks the support from Fundação para a Ciência e a Tecnologia (Portugal), namely through the project UID/EEA/50008/2013.

 \end{document}